\documentclass[reprint, amsmath, amssymb, aps, prx]{revtex4-2}

\usepackage[ruled,lined]{algorithm2e}

\usepackage{bm}
\usepackage{bbm}
\usepackage{braket}
\usepackage{dcolumn}
\usepackage{dsfont}
\usepackage{graphicx}
\usepackage{makecell}
\usepackage{nicefrac}
\usepackage{pifont}
\usepackage{placeins}
\usepackage{stmaryrd}
\usepackage[caption=false]{subfig}

\usepackage{pgfkeys}

\usepackage{tikz}
\usetikzlibrary{calc}
\usetikzlibrary{math}

\usepackage{hyperref}
\usepackage[noabbrev,capitalise]{cleveref}

\newcommand{\cnot}{\textsc{CNot}}
\renewcommand{\set}[1]{\left \lbrace #1 \right \rbrace}
\newcommand{\com}[2]{\left[ #1,\, #2 \right]}
\newcommand{\acom}[2]{\left \lbrace #1,\, #2 \right \rbrace}
\newcommand{\abs}[1]{\left \lvert #1 \right \rvert}
\newcommand{\nkd}[3]{\left \llbracket #1, \, #2,\, #3 \right \rrbracket}

\DeclareMathOperator*{\argmin}{arg\,min}

\begin{document}

\preprint{APS/123-QED}

\title{Automated Flag-based Fault-Tolerant State Preparation using Integer Linear Programming}

\author{Ben Criger}
\email{ben.criger@quantinuum.com}
\affiliation{Quantinuum, Terrington House, 13-15 Hills Road, Cambridge CB2 1NL, UK}\author{Aaron Hankin}
\author{Justin J. Burau}
\author{Abigail R. Perry}
\affiliation{Quantinuum, 303 South Technology Court, Broomfield, Colorado 80021, USA}

\date{\today}

\begin{abstract}
Post-selected stabilizer state preparation is a necessary subroutine in fault-tolerant quantum computation, both for initialization of logical qubits, and for logical-ancilla-based error correction gadgets (e.g. Steane and Knill).
Therefore, reducing the number of gates needed to prepare a stabilizer state fault-tolerantly can simultaneously reduce time-to-solution and increase reliability. 
For small, low-distance codes such as the $\nkd{7}{1}{3}$ Steane code, circuits with low gate counts can be found by inspection. 
This becomes impractical for larger codes, necessitating automation. 
There are two state-of-the-art methods for automated fault-tolerant state preparation, SAT-based stabilizer measurement and flag-at-origin. 
In this work, we optimize state preparation circuits using the circuit gauge operator formalism to express the construction of flag circuits as an integer linear program. 
This allows the construction of circuits with equal or lower gate count than the state of the art, while detecting up to three errors. 
We use this technique to derive a Steane error correction gadget for the $\nkd{24}{10}{4}$ two-block group algebra code, and test it on Quantinuum's System Model H2 quantum computer with $10^4$ shots, resulting in a logical block error rate $\sim 1.4 \times 10^{-4}$ ($\sim 1.4 \times 10^{-5}$ per logical qubit), with $\sim 1.6\%$ of the shots post-selected due to weight-two errors. 
\end{abstract}

\maketitle

\section{Introduction}
\label{sec:intro}
Quantum computers have a large and growing number of potential applications \cite{montanaroQuantumAlgorithmsOverview2016}, and quantum computers of increasing size are being constructed \cite{ransford2025helios98qubittrappedionquantum,GoogleWillow}.
Owing to the effects of noise and physical imperfections, these devices continue to have physical error rates on the order of $8 \times 10^{-4}$ \cite{ransford2025helios98qubittrappedionquantum} for a typical two-qubit entangling gate or measurement, preventing the direct implementation of large-scale algorithms with physical qubits.
Therefore, it is necessary to carry out a quantum computation fault-tolerantly, interleaving logical operations with quantum error correction subroutines (called \emph{QEC gadgets}).

QEC gadget design has been a critical area for study since the inception of QEC. 
This is because the logical error rate for a QEC gadget with $g$ gates, capable of tolerating $t$ faults that each occur with probability $p$ is, to a first approximation
\begin{equation}
\alpha(g, t) {g \choose t + 1} p^{t + 1},
\end{equation}
where $\alpha(g, t)$ is the fraction of fault sets of size $t + 1$ that result in logical error\footnote{We note that, while more sophisticated approximations of logical error rates have recently become available \cite{IBMLERModel}, the crude model used above still suffices for qualitative analysis}.
This implies that a reduction in gate count for a QEC gadget will result in a larger reduction in logical error rate, and that this effect will become more pronounced when using gadgets designed to tolerate higher numbers of faults. 
Additionally, the presence of time-dependent coherent errors in QCCD devices implies that a reduction in gate count will also reduce \emph{physical} error rates directly. 

Reliability of subroutines in a quantum computation can also be improved through post-selection, at the cost of increasing the number of times a computation must be repeated. 
In the context of QEC, this overhead increase can be reduced by using Knill \cite{KnillEC} or Steane \cite{SteaneEC} error correction circuits, both of which effectively reduce the task of QEC to state preparation, which can be attempted using physical qubits which are separate from those being used for computation. 
This allows state preparation to be repeated in case of failure, without requiring an entire computation to be repeated.

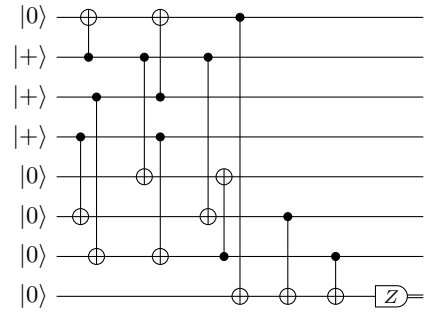
\begin{figure}[!htbp]
\centering
\begin{tikzpicture}[scale=1.000000,x=1pt,y=1pt]
\filldraw[color=white] (0.000000, -7.500000) rectangle (138.000000, 112.500000);
% Drawing wires
% Line 1: 0 W \ket{0}
\draw[color=black] (0.000000,105.000000) -- (138.000000,105.000000);
\draw[color=black] (0.000000,105.000000) node[left] {$\ket{0}$};
% Line 2: 1 W \ket{+}
\draw[color=black] (0.000000,90.000000) -- (138.000000,90.000000);
\draw[color=black] (0.000000,90.000000) node[left] {$\ket{+}$};
% Line 3: 2 W \ket{+}
\draw[color=black] (0.000000,75.000000) -- (138.000000,75.000000);
\draw[color=black] (0.000000,75.000000) node[left] {$\ket{+}$};
% Line 4: 3 W \ket{+}
\draw[color=black] (0.000000,60.000000) -- (138.000000,60.000000);
\draw[color=black] (0.000000,60.000000) node[left] {$\ket{+}$};
% Line 5: 4 W \ket{0}
\draw[color=black] (0.000000,45.000000) -- (138.000000,45.000000);
\draw[color=black] (0.000000,45.000000) node[left] {$\ket{0}$};
% Line 6: 5 W \ket{0}
\draw[color=black] (0.000000,30.000000) -- (138.000000,30.000000);
\draw[color=black] (0.000000,30.000000) node[left] {$\ket{0}$};
% Line 7: 6 W \ket{0}
\draw[color=black] (0.000000,15.000000) -- (138.000000,15.000000);
\draw[color=black] (0.000000,15.000000) node[left] {$\ket{0}$};
% Line 8: A W \ket{0}
\draw[color=black] (0.000000,0.000000) -- (126.000000,0.000000);
\draw[color=black] (126.000000,-0.500000) -- (138.000000,-0.500000);
\draw[color=black] (126.000000,0.500000) -- (138.000000,0.500000);
\draw[color=black] (0.000000,0.000000) node[left] {$\ket{0}$};
% Done with wires; drawing gates
% Line 10: +0 1
\draw (12.000000,105.000000) -- (12.000000,90.000000);
\begin{scope}
\draw[fill=white] (12.000000, 105.000000) circle(3.000000pt);
\clip (12.000000, 105.000000) circle(3.000000pt);
\draw (9.000000, 105.000000) -- (15.000000, 105.000000);
\draw (12.000000, 102.000000) -- (12.000000, 108.000000);
\end{scope}
\filldraw (12.000000, 90.000000) circle(1.500000pt);
% Line 11: +5 3
\draw (9.000000,60.000000) -- (9.000000,30.000000);
\begin{scope}
\draw[fill=white] (9.000000, 30.000000) circle(3.000000pt);
\clip (9.000000, 30.000000) circle(3.000000pt);
\draw (6.000000, 30.000000) -- (12.000000, 30.000000);
\draw (9.000000, 27.000000) -- (9.000000, 33.000000);
\end{scope}
\filldraw (9.000000, 60.000000) circle(1.500000pt);
% Line 12: +6 2
\draw (15.000000,75.000000) -- (15.000000,15.000000);
\begin{scope}
\draw[fill=white] (15.000000, 15.000000) circle(3.000000pt);
\clip (15.000000, 15.000000) circle(3.000000pt);
\draw (12.000000, 15.000000) -- (18.000000, 15.000000);
\draw (15.000000, 12.000000) -- (15.000000, 18.000000);
\end{scope}
\filldraw (15.000000, 75.000000) circle(1.500000pt);
% Line 13: +4 1
\draw (33.000000,90.000000) -- (33.000000,45.000000);
\begin{scope}
\draw[fill=white] (33.000000, 45.000000) circle(3.000000pt);
\clip (33.000000, 45.000000) circle(3.000000pt);
\draw (30.000000, 45.000000) -- (36.000000, 45.000000);
\draw (33.000000, 42.000000) -- (33.000000, 48.000000);
\end{scope}
\filldraw (33.000000, 90.000000) circle(1.500000pt);
% Line 14: +0 2
\draw (39.000000,105.000000) -- (39.000000,75.000000);
\begin{scope}
\draw[fill=white] (39.000000, 105.000000) circle(3.000000pt);
\clip (39.000000, 105.000000) circle(3.000000pt);
\draw (36.000000, 105.000000) -- (42.000000, 105.000000);
\draw (39.000000, 102.000000) -- (39.000000, 108.000000);
\end{scope}
\filldraw (39.000000, 75.000000) circle(1.500000pt);
% Line 15: +6 3
\draw (39.000000,60.000000) -- (39.000000,15.000000);
\begin{scope}
\draw[fill=white] (39.000000, 15.000000) circle(3.000000pt);
\clip (39.000000, 15.000000) circle(3.000000pt);
\draw (36.000000, 15.000000) -- (42.000000, 15.000000);
\draw (39.000000, 12.000000) -- (39.000000, 18.000000);
\end{scope}
\filldraw (39.000000, 60.000000) circle(1.500000pt);
% Line 16: +5 1
\draw (57.000000,90.000000) -- (57.000000,30.000000);
\begin{scope}
\draw[fill=white] (57.000000, 30.000000) circle(3.000000pt);
\clip (57.000000, 30.000000) circle(3.000000pt);
\draw (54.000000, 30.000000) -- (60.000000, 30.000000);
\draw (57.000000, 27.000000) -- (57.000000, 33.000000);
\end{scope}
\filldraw (57.000000, 90.000000) circle(1.500000pt);
% Line 17: +4 6
\draw (63.000000,45.000000) -- (63.000000,15.000000);
\begin{scope}
\draw[fill=white] (63.000000, 45.000000) circle(3.000000pt);
\clip (63.000000, 45.000000) circle(3.000000pt);
\draw (60.000000, 45.000000) -- (66.000000, 45.000000);
\draw (63.000000, 42.000000) -- (63.000000, 48.000000);
\end{scope}
\filldraw (63.000000, 15.000000) circle(1.500000pt);
% Line 18: +A 0
\draw (69.000000,105.000000) -- (69.000000,0.000000);
\begin{scope}
\draw[fill=white] (69.000000, 0.000000) circle(3.000000pt);
\clip (69.000000, 0.000000) circle(3.000000pt);
\draw (66.000000, 0.000000) -- (72.000000, 0.000000);
\draw (69.000000, -3.000000) -- (69.000000, 3.000000);
\end{scope}
\filldraw (69.000000, 105.000000) circle(1.500000pt);
% Line 19: +A 5
\draw (87.000000,30.000000) -- (87.000000,0.000000);
\begin{scope}
\draw[fill=white] (87.000000, 0.000000) circle(3.000000pt);
\clip (87.000000, 0.000000) circle(3.000000pt);
\draw (84.000000, 0.000000) -- (90.000000, 0.000000);
\draw (87.000000, -3.000000) -- (87.000000, 3.000000);
\end{scope}
\filldraw (87.000000, 30.000000) circle(1.500000pt);
% Line 20: +A 6
\draw (105.000000,15.000000) -- (105.000000,0.000000);
\begin{scope}
\draw[fill=white] (105.000000, 0.000000) circle(3.000000pt);
\clip (105.000000, 0.000000) circle(3.000000pt);
\draw (102.000000, 0.000000) -- (108.000000, 0.000000);
\draw (105.000000, -3.000000) -- (105.000000, 3.000000);
\end{scope}
\filldraw (105.000000, 15.000000) circle(1.500000pt);
% Line 21: A M {\scriptsize{$Z$}}
\draw[fill=white] (120.000000, -4.000000) -- (128.000000,-4.000000) arc (-90:90:4.000000pt) -- (120.000000,4.000000) -- cycle;
\draw (126.000000, 0.000000) node {{\scriptsize{$Z$}}};
% Done with gates; drawing ending labels
% Done with ending labels; drawing cut lines and comments
% Done with comments
\end{tikzpicture}
\caption{From \cite{GotoSteane}; a circuit for fault-tolerant $\ket{\overline{0}}$ state preparation for the $\nkd{7}{1}{3}$ Steane code, incorporating measurement of a single weight-3 operator to achieve a total operation count of 12.}
\label{fig:GotoOriginalCircuit}
\end{figure}

For these reasons, the post-selected preparation of encoded states can be a critical step in QEC. 
An archetypal example of fault-tolerant state preparation was derived by Goto \cite{GotoSteane}, see \Cref{fig:GotoOriginalCircuit}.
This state preparation circuit begins with a non-fault-tolerant stage, which is followed by measurement of a single weight-3 operator, requiring twelve high-error-rate physical operations (i.e. two-qubit gates and measurements), far fewer than previous methods.
However, the circuit was derived by inspection, and no algorithm for circuit derivation was presented. 

Since Goto's original work, two procedures to automate the derivation of state preparation circuits have been devised. 
Peham, et al \cite{PehamSAT} iteratively increase the number of faults a circuit can tolerate (a process we refer to as \emph{promotion} in the remainder of this work) using a SAT solver to derive a set of low-weight operators that anticommute with high-weight errors output from the circuit under construction, then measure these operators with fault-tolerant flag circuits. 
Forlivesi and Amaro \cite{FlagAtOrigin} automate the construction of flag-based circuits \cite{ReichardtFlags} for state preparation in polynomial time, using a pre-constructed library of subcircuits, without any optimization. 

In this work, we use the formalism of circuit gauge operators \cite{CircuitGaugeOperators} to solve the promotion problem, combining stabilizer measurements and flag subcircuits at each step. 
We begin with a brief presentation of circuit gauge operators in \Cref{sec:CircuitGaugeOperators}, including the formalism necessary to derive the main theoretical results. 
We then reduce the promotion problem to an integer linear program \cite{IntegerLinearPrograms} in \Cref{sec:MapToIntegerLinearProgram},
show concrete examples for selected CSS codes in \Cref{sec:Examples}, compare error rates for various state preparation circuits in simulation and hardware in \Cref{sec:Results}, and conclude in \Cref{sec:DiscussionAndConclusion}.

\section{Circuit Gauge Operators}
\label{sec:CircuitGaugeOperators}

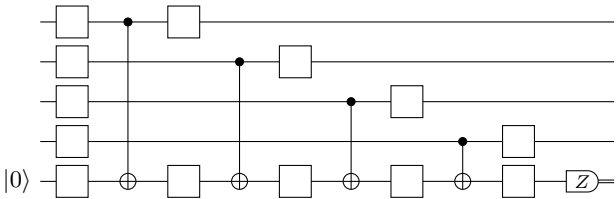
\begin{figure}[htbp!]
\centering
\begin{tikzpicture}[scale=1.000000,x=1pt,y=1pt]
\filldraw[color=white] (0.000000, -7.500000) rectangle (216.000000, 67.500000);
% Drawing wires
% Line 1: 1 W
\draw[color=black] (0.000000,60.000000) -- (216.000000,60.000000);
% Line 2: 2 W
\draw[color=black] (0.000000,45.000000) -- (216.000000,45.000000);
% Line 3: 3 W
\draw[color=black] (0.000000,30.000000) -- (216.000000,30.000000);
% Line 4: 4 W
\draw[color=black] (0.000000,15.000000) -- (216.000000,15.000000);
% Line 5: A W \ket{0}
\draw[color=black] (0.000000,0.000000) -- (204.000000,0.000000);
\draw[color=black] (204.000000,-0.500000) -- (216.000000,-0.500000);
\draw[color=black] (204.000000,0.500000) -- (216.000000,0.500000);
\draw[color=black] (0.000000,0.000000) node[left] {$\ket{0}$};
% Done with wires; drawing gates
% Line 7: 1 G {}
\begin{scope}
\draw[fill=white] (12.000000, 60.000000) +(-45.000000:8.485281pt and 8.485281pt) -- +(45.000000:8.485281pt and 8.485281pt) -- +(135.000000:8.485281pt and 8.485281pt) -- +(225.000000:8.485281pt and 8.485281pt) -- cycle;
\clip (12.000000, 60.000000) +(-45.000000:8.485281pt and 8.485281pt) -- +(45.000000:8.485281pt and 8.485281pt) -- +(135.000000:8.485281pt and 8.485281pt) -- +(225.000000:8.485281pt and 8.485281pt) -- cycle;
\draw (12.000000, 60.000000) node {{}};
\end{scope}
% Line 8: 2 G {}
\begin{scope}
\draw[fill=white] (12.000000, 45.000000) +(-45.000000:8.485281pt and 8.485281pt) -- +(45.000000:8.485281pt and 8.485281pt) -- +(135.000000:8.485281pt and 8.485281pt) -- +(225.000000:8.485281pt and 8.485281pt) -- cycle;
\clip (12.000000, 45.000000) +(-45.000000:8.485281pt and 8.485281pt) -- +(45.000000:8.485281pt and 8.485281pt) -- +(135.000000:8.485281pt and 8.485281pt) -- +(225.000000:8.485281pt and 8.485281pt) -- cycle;
\draw (12.000000, 45.000000) node {{}};
\end{scope}
% Line 9: 3 G {}
\begin{scope}
\draw[fill=white] (12.000000, 30.000000) +(-45.000000:8.485281pt and 8.485281pt) -- +(45.000000:8.485281pt and 8.485281pt) -- +(135.000000:8.485281pt and 8.485281pt) -- +(225.000000:8.485281pt and 8.485281pt) -- cycle;
\clip (12.000000, 30.000000) +(-45.000000:8.485281pt and 8.485281pt) -- +(45.000000:8.485281pt and 8.485281pt) -- +(135.000000:8.485281pt and 8.485281pt) -- +(225.000000:8.485281pt and 8.485281pt) -- cycle;
\draw (12.000000, 30.000000) node {{}};
\end{scope}
% Line 10: 4 G {}
\begin{scope}
\draw[fill=white] (12.000000, 15.000000) +(-45.000000:8.485281pt and 8.485281pt) -- +(45.000000:8.485281pt and 8.485281pt) -- +(135.000000:8.485281pt and 8.485281pt) -- +(225.000000:8.485281pt and 8.485281pt) -- cycle;
\clip (12.000000, 15.000000) +(-45.000000:8.485281pt and 8.485281pt) -- +(45.000000:8.485281pt and 8.485281pt) -- +(135.000000:8.485281pt and 8.485281pt) -- +(225.000000:8.485281pt and 8.485281pt) -- cycle;
\draw (12.000000, 15.000000) node {{}};
\end{scope}
% Line 12: A G {}
\begin{scope}
\draw[fill=white] (12.000000, -0.000000) +(-45.000000:8.485281pt and 8.485281pt) -- +(45.000000:8.485281pt and 8.485281pt) -- +(135.000000:8.485281pt and 8.485281pt) -- +(225.000000:8.485281pt and 8.485281pt) -- cycle;
\clip (12.000000, -0.000000) +(-45.000000:8.485281pt and 8.485281pt) -- +(45.000000:8.485281pt and 8.485281pt) -- +(135.000000:8.485281pt and 8.485281pt) -- +(225.000000:8.485281pt and 8.485281pt) -- cycle;
\draw (12.000000, -0.000000) node {{}};
\end{scope}
% Line 14: +A 1
\draw (33.000000,60.000000) -- (33.000000,0.000000);
\begin{scope}
\draw[fill=white] (33.000000, 0.000000) circle(3.000000pt);
\clip (33.000000, 0.000000) circle(3.000000pt);
\draw (30.000000, 0.000000) -- (36.000000, 0.000000);
\draw (33.000000, -3.000000) -- (33.000000, 3.000000);
\end{scope}
\filldraw (33.000000, 60.000000) circle(1.500000pt);
% Line 15: A G {}
\begin{scope}
\draw[fill=white] (54.000000, -0.000000) +(-45.000000:8.485281pt and 8.485281pt) -- +(45.000000:8.485281pt and 8.485281pt) -- +(135.000000:8.485281pt and 8.485281pt) -- +(225.000000:8.485281pt and 8.485281pt) -- cycle;
\clip (54.000000, -0.000000) +(-45.000000:8.485281pt and 8.485281pt) -- +(45.000000:8.485281pt and 8.485281pt) -- +(135.000000:8.485281pt and 8.485281pt) -- +(225.000000:8.485281pt and 8.485281pt) -- cycle;
\draw (54.000000, -0.000000) node {{}};
\end{scope}
% Line 16: 1 G {}
\begin{scope}
\draw[fill=white] (54.000000, 60.000000) +(-45.000000:8.485281pt and 8.485281pt) -- +(45.000000:8.485281pt and 8.485281pt) -- +(135.000000:8.485281pt and 8.485281pt) -- +(225.000000:8.485281pt and 8.485281pt) -- cycle;
\clip (54.000000, 60.000000) +(-45.000000:8.485281pt and 8.485281pt) -- +(45.000000:8.485281pt and 8.485281pt) -- +(135.000000:8.485281pt and 8.485281pt) -- +(225.000000:8.485281pt and 8.485281pt) -- cycle;
\draw (54.000000, 60.000000) node {{}};
\end{scope}
% Line 18: +A 2
\draw (75.000000,45.000000) -- (75.000000,0.000000);
\begin{scope}
\draw[fill=white] (75.000000, 0.000000) circle(3.000000pt);
\clip (75.000000, 0.000000) circle(3.000000pt);
\draw (72.000000, 0.000000) -- (78.000000, 0.000000);
\draw (75.000000, -3.000000) -- (75.000000, 3.000000);
\end{scope}
\filldraw (75.000000, 45.000000) circle(1.500000pt);
% Line 19: A G {}
\begin{scope}
\draw[fill=white] (96.000000, -0.000000) +(-45.000000:8.485281pt and 8.485281pt) -- +(45.000000:8.485281pt and 8.485281pt) -- +(135.000000:8.485281pt and 8.485281pt) -- +(225.000000:8.485281pt and 8.485281pt) -- cycle;
\clip (96.000000, -0.000000) +(-45.000000:8.485281pt and 8.485281pt) -- +(45.000000:8.485281pt and 8.485281pt) -- +(135.000000:8.485281pt and 8.485281pt) -- +(225.000000:8.485281pt and 8.485281pt) -- cycle;
\draw (96.000000, -0.000000) node {{}};
\end{scope}
% Line 20: 2 G {}
\begin{scope}
\draw[fill=white] (96.000000, 45.000000) +(-45.000000:8.485281pt and 8.485281pt) -- +(45.000000:8.485281pt and 8.485281pt) -- +(135.000000:8.485281pt and 8.485281pt) -- +(225.000000:8.485281pt and 8.485281pt) -- cycle;
\clip (96.000000, 45.000000) +(-45.000000:8.485281pt and 8.485281pt) -- +(45.000000:8.485281pt and 8.485281pt) -- +(135.000000:8.485281pt and 8.485281pt) -- +(225.000000:8.485281pt and 8.485281pt) -- cycle;
\draw (96.000000, 45.000000) node {{}};
\end{scope}
% Line 22: +A 3
\draw (117.000000,30.000000) -- (117.000000,0.000000);
\begin{scope}
\draw[fill=white] (117.000000, 0.000000) circle(3.000000pt);
\clip (117.000000, 0.000000) circle(3.000000pt);
\draw (114.000000, 0.000000) -- (120.000000, 0.000000);
\draw (117.000000, -3.000000) -- (117.000000, 3.000000);
\end{scope}
\filldraw (117.000000, 30.000000) circle(1.500000pt);
% Line 23: A G {}
\begin{scope}
\draw[fill=white] (138.000000, -0.000000) +(-45.000000:8.485281pt and 8.485281pt) -- +(45.000000:8.485281pt and 8.485281pt) -- +(135.000000:8.485281pt and 8.485281pt) -- +(225.000000:8.485281pt and 8.485281pt) -- cycle;
\clip (138.000000, -0.000000) +(-45.000000:8.485281pt and 8.485281pt) -- +(45.000000:8.485281pt and 8.485281pt) -- +(135.000000:8.485281pt and 8.485281pt) -- +(225.000000:8.485281pt and 8.485281pt) -- cycle;
\draw (138.000000, -0.000000) node {{}};
\end{scope}
% Line 24: 3 G {}
\begin{scope}
\draw[fill=white] (138.000000, 30.000000) +(-45.000000:8.485281pt and 8.485281pt) -- +(45.000000:8.485281pt and 8.485281pt) -- +(135.000000:8.485281pt and 8.485281pt) -- +(225.000000:8.485281pt and 8.485281pt) -- cycle;
\clip (138.000000, 30.000000) +(-45.000000:8.485281pt and 8.485281pt) -- +(45.000000:8.485281pt and 8.485281pt) -- +(135.000000:8.485281pt and 8.485281pt) -- +(225.000000:8.485281pt and 8.485281pt) -- cycle;
\draw (138.000000, 30.000000) node {{}};
\end{scope}
% Line 26: +A 4
\draw (159.000000,15.000000) -- (159.000000,0.000000);
\begin{scope}
\draw[fill=white] (159.000000, 0.000000) circle(3.000000pt);
\clip (159.000000, 0.000000) circle(3.000000pt);
\draw (156.000000, 0.000000) -- (162.000000, 0.000000);
\draw (159.000000, -3.000000) -- (159.000000, 3.000000);
\end{scope}
\filldraw (159.000000, 15.000000) circle(1.500000pt);
% Line 27: A G {}
\begin{scope}
\draw[fill=white] (180.000000, -0.000000) +(-45.000000:8.485281pt and 8.485281pt) -- +(45.000000:8.485281pt and 8.485281pt) -- +(135.000000:8.485281pt and 8.485281pt) -- +(225.000000:8.485281pt and 8.485281pt) -- cycle;
\clip (180.000000, -0.000000) +(-45.000000:8.485281pt and 8.485281pt) -- +(45.000000:8.485281pt and 8.485281pt) -- +(135.000000:8.485281pt and 8.485281pt) -- +(225.000000:8.485281pt and 8.485281pt) -- cycle;
\draw (180.000000, -0.000000) node {{}};
\end{scope}
% Line 28: 4 G {}
\begin{scope}
\draw[fill=white] (180.000000, 15.000000) +(-45.000000:8.485281pt and 8.485281pt) -- +(45.000000:8.485281pt and 8.485281pt) -- +(135.000000:8.485281pt and 8.485281pt) -- +(225.000000:8.485281pt and 8.485281pt) -- cycle;
\clip (180.000000, 15.000000) +(-45.000000:8.485281pt and 8.485281pt) -- +(45.000000:8.485281pt and 8.485281pt) -- +(135.000000:8.485281pt and 8.485281pt) -- +(225.000000:8.485281pt and 8.485281pt) -- cycle;
\draw (180.000000, 15.000000) node {{}};
\end{scope}
% Line 30: A M {\scriptsize{$Z$}}
\draw[fill=white] (198.000000, -4.000000) -- (206.000000,-4.000000) arc (-90:90:4.000000pt) -- (198.000000,4.000000) -- cycle;
\draw (204.000000, 0.000000) node {{\scriptsize{$Z$}}};
% Done with gates; drawing ending labels
% Done with ending labels; drawing cut lines and comments
% Done with comments
\end{tikzpicture}
\caption{A subcircuit measuring $Z^{\otimes 4}$, with locations represented using empty boxes.
Note that this set of locations, while capable of describing the support of any fault caused by a faulty operation, does not have support on every qubit at every point in time, as in \cite{DelfosseGaugeOperators}.}
\label{fig:CircuitWithLocations}
\end{figure}

In order to automate the construction of fault-tolerant circuits that include flag sub-circuits, we express them as measurements of Pauli operators. 
These Pauli operators, rather than being supported on the qubits at the output of the circuit, are supported on the \emph{locations} within the circuit.
A location within a circuit is described by a tuple $(q,\, t)$ corresponding to qubit $q$ and time $t$, at which an error may be supported. 
Throughout this work, we place these locations on uninitialized qubits at the beginning of the circuit, and after each physical preparation and two-qubit gate (none of the circuits used in this work require one-qubit gates), see \Cref{fig:CircuitWithLocations}. 

\begin{figure}[htbp!]
\centering
\begin{tikzpicture}
\node at (0, 0.5) {\begin{tikzpicture}[scale=1.000000,x=1pt,y=1pt]
\filldraw[color=white] (0.000000, -7.500000) rectangle (24.000000, 7.500000);
% Drawing wires
% Line 1: 0 W \ket{0}
\draw[color=black] (0.000000,0.000000) -- (24.000000,0.000000);
\draw[color=black] (0.000000,0.000000) node[left] {$\ket{0}$};
% Done with wires; drawing gates
% Line 3: 0 G $Z$
\begin{scope}
\draw[fill=white] (12.000000, -0.000000) +(-45.000000:8.485281pt and 8.485281pt) -- +(45.000000:8.485281pt and 8.485281pt) -- +(135.000000:8.485281pt and 8.485281pt) -- +(225.000000:8.485281pt and 8.485281pt) -- cycle;
\clip (12.000000, -0.000000) +(-45.000000:8.485281pt and 8.485281pt) -- +(45.000000:8.485281pt and 8.485281pt) -- +(135.000000:8.485281pt and 8.485281pt) -- +(225.000000:8.485281pt and 8.485281pt) -- cycle;
\draw (12.000000, -0.000000) node {$Z$};
\end{scope}
% Done with gates; drawing ending labels
% Done with ending labels; drawing cut lines and comments
% Done with comments
\end{tikzpicture}};
\node at (0, -0.5) {\begin{tikzpicture}[scale=1.000000,x=1pt,y=1pt]
\filldraw[color=white] (0.000000, -7.500000) rectangle (24.000000, 7.500000);
% Drawing wires
% Line 1: 0 W \ket{+}
\draw[color=black] (0.000000,0.000000) -- (24.000000,0.000000);
\draw[color=black] (0.000000,0.000000) node[left] {$\ket{+}$};
% Done with wires; drawing gates
% Line 3: 0 G $X$
\begin{scope}
\draw[fill=white] (12.000000, -0.000000) +(-45.000000:8.485281pt and 8.485281pt) -- +(45.000000:8.485281pt and 8.485281pt) -- +(135.000000:8.485281pt and 8.485281pt) -- +(225.000000:8.485281pt and 8.485281pt) -- cycle;
\clip (12.000000, -0.000000) +(-45.000000:8.485281pt and 8.485281pt) -- +(45.000000:8.485281pt and 8.485281pt) -- +(135.000000:8.485281pt and 8.485281pt) -- +(225.000000:8.485281pt and 8.485281pt) -- cycle;
\draw (12.000000, -0.000000) node {$X$};
\end{scope}
% Done with gates; drawing ending labels
% Done with ending labels; drawing cut lines and comments
% Done with comments
\end{tikzpicture}};

\node at (2.5, 1.8) {\begin{tikzpicture}[scale=1.000000,x=1pt,y=1pt]
\filldraw[color=white] (0.000000, -7.500000) rectangle (66.000000, 22.500000);
% Drawing wires
% Line 1: 0 W
\draw[color=black] (0.000000,15.000000) -- (66.000000,15.000000);
% Line 2: 1 W
\draw[color=black] (0.000000,0.000000) -- (66.000000,0.000000);
% Done with wires; drawing gates
% Line 4: 0 G {$X$}
\begin{scope}
\draw[fill=white] (12.000000, 15.000000) +(-45.000000:8.485281pt and 8.485281pt) -- +(45.000000:8.485281pt and 8.485281pt) -- +(135.000000:8.485281pt and 8.485281pt) -- +(225.000000:8.485281pt and 8.485281pt) -- cycle;
\clip (12.000000, 15.000000) +(-45.000000:8.485281pt and 8.485281pt) -- +(45.000000:8.485281pt and 8.485281pt) -- +(135.000000:8.485281pt and 8.485281pt) -- +(225.000000:8.485281pt and 8.485281pt) -- cycle;
\draw (12.000000, 15.000000) node {{$X$}};
\end{scope}
% Line 5: 1 G {}
\begin{scope}
\draw[fill=white] (12.000000, -0.000000) +(-45.000000:8.485281pt and 8.485281pt) -- +(45.000000:8.485281pt and 8.485281pt) -- +(135.000000:8.485281pt and 8.485281pt) -- +(225.000000:8.485281pt and 8.485281pt) -- cycle;
\clip (12.000000, -0.000000) +(-45.000000:8.485281pt and 8.485281pt) -- +(45.000000:8.485281pt and 8.485281pt) -- +(135.000000:8.485281pt and 8.485281pt) -- +(225.000000:8.485281pt and 8.485281pt) -- cycle;
\draw (12.000000, -0.000000) node {{}};
\end{scope}
% Line 6: +1 0
\draw (33.000000,15.000000) -- (33.000000,0.000000);
\begin{scope}
\draw[fill=white] (33.000000, 0.000000) circle(3.000000pt);
\clip (33.000000, 0.000000) circle(3.000000pt);
\draw (30.000000, 0.000000) -- (36.000000, 0.000000);
\draw (33.000000, -3.000000) -- (33.000000, 3.000000);
\end{scope}
\filldraw (33.000000, 15.000000) circle(1.500000pt);
% Line 7: 0 G {$X$}
\begin{scope}
\draw[fill=white] (54.000000, 15.000000) +(-45.000000:8.485281pt and 8.485281pt) -- +(45.000000:8.485281pt and 8.485281pt) -- +(135.000000:8.485281pt and 8.485281pt) -- +(225.000000:8.485281pt and 8.485281pt) -- cycle;
\clip (54.000000, 15.000000) +(-45.000000:8.485281pt and 8.485281pt) -- +(45.000000:8.485281pt and 8.485281pt) -- +(135.000000:8.485281pt and 8.485281pt) -- +(225.000000:8.485281pt and 8.485281pt) -- cycle;
\draw (54.000000, 15.000000) node {{$X$}};
\end{scope}
% Line 8: 1 G {$X$}
\begin{scope}
\draw[fill=white] (54.000000, -0.000000) +(-45.000000:8.485281pt and 8.485281pt) -- +(45.000000:8.485281pt and 8.485281pt) -- +(135.000000:8.485281pt and 8.485281pt) -- +(225.000000:8.485281pt and 8.485281pt) -- cycle;
\clip (54.000000, -0.000000) +(-45.000000:8.485281pt and 8.485281pt) -- +(45.000000:8.485281pt and 8.485281pt) -- +(135.000000:8.485281pt and 8.485281pt) -- +(225.000000:8.485281pt and 8.485281pt) -- cycle;
\draw (54.000000, -0.000000) node {{$X$}};
\end{scope}
% Done with gates; drawing ending labels
% Done with ending labels; drawing cut lines and comments
% Done with comments
\end{tikzpicture}};
\node at (2.5, 0.6) {\begin{tikzpicture}[scale=1.000000,x=1pt,y=1pt]
\filldraw[color=white] (0.000000, -7.500000) rectangle (66.000000, 22.500000);
% Drawing wires
% Line 1: 0 W
\draw[color=black] (0.000000,15.000000) -- (66.000000,15.000000);
% Line 2: 1 W
\draw[color=black] (0.000000,0.000000) -- (66.000000,0.000000);
% Done with wires; drawing gates
% Line 4: 0 G {}
\begin{scope}
\draw[fill=white] (12.000000, 15.000000) +(-45.000000:8.485281pt and 8.485281pt) -- +(45.000000:8.485281pt and 8.485281pt) -- +(135.000000:8.485281pt and 8.485281pt) -- +(225.000000:8.485281pt and 8.485281pt) -- cycle;
\clip (12.000000, 15.000000) +(-45.000000:8.485281pt and 8.485281pt) -- +(45.000000:8.485281pt and 8.485281pt) -- +(135.000000:8.485281pt and 8.485281pt) -- +(225.000000:8.485281pt and 8.485281pt) -- cycle;
\draw (12.000000, 15.000000) node {{}};
\end{scope}
% Line 5: 1 G {$Z$}
\begin{scope}
\draw[fill=white] (12.000000, -0.000000) +(-45.000000:8.485281pt and 8.485281pt) -- +(45.000000:8.485281pt and 8.485281pt) -- +(135.000000:8.485281pt and 8.485281pt) -- +(225.000000:8.485281pt and 8.485281pt) -- cycle;
\clip (12.000000, -0.000000) +(-45.000000:8.485281pt and 8.485281pt) -- +(45.000000:8.485281pt and 8.485281pt) -- +(135.000000:8.485281pt and 8.485281pt) -- +(225.000000:8.485281pt and 8.485281pt) -- cycle;
\draw (12.000000, -0.000000) node {{$Z$}};
\end{scope}
% Line 6: +1 0
\draw (33.000000,15.000000) -- (33.000000,0.000000);
\begin{scope}
\draw[fill=white] (33.000000, 0.000000) circle(3.000000pt);
\clip (33.000000, 0.000000) circle(3.000000pt);
\draw (30.000000, 0.000000) -- (36.000000, 0.000000);
\draw (33.000000, -3.000000) -- (33.000000, 3.000000);
\end{scope}
\filldraw (33.000000, 15.000000) circle(1.500000pt);
% Line 7: 0 G {$Z$}
\begin{scope}
\draw[fill=white] (54.000000, 15.000000) +(-45.000000:8.485281pt and 8.485281pt) -- +(45.000000:8.485281pt and 8.485281pt) -- +(135.000000:8.485281pt and 8.485281pt) -- +(225.000000:8.485281pt and 8.485281pt) -- cycle;
\clip (54.000000, 15.000000) +(-45.000000:8.485281pt and 8.485281pt) -- +(45.000000:8.485281pt and 8.485281pt) -- +(135.000000:8.485281pt and 8.485281pt) -- +(225.000000:8.485281pt and 8.485281pt) -- cycle;
\draw (54.000000, 15.000000) node {{$Z$}};
\end{scope}
% Line 8: 1 G {$Z$}
\begin{scope}
\draw[fill=white] (54.000000, -0.000000) +(-45.000000:8.485281pt and 8.485281pt) -- +(45.000000:8.485281pt and 8.485281pt) -- +(135.000000:8.485281pt and 8.485281pt) -- +(225.000000:8.485281pt and 8.485281pt) -- cycle;
\clip (54.000000, -0.000000) +(-45.000000:8.485281pt and 8.485281pt) -- +(45.000000:8.485281pt and 8.485281pt) -- +(135.000000:8.485281pt and 8.485281pt) -- +(225.000000:8.485281pt and 8.485281pt) -- cycle;
\draw (54.000000, -0.000000) node {{$Z$}};
\end{scope}
% Done with gates; drawing ending labels
% Done with ending labels; drawing cut lines and comments
% Done with comments
\end{tikzpicture}};
\node at (2.5, -0.6) {\begin{tikzpicture}[scale=1.000000,x=1pt,y=1pt]
\filldraw[color=white] (0.000000, -7.500000) rectangle (66.000000, 22.500000);
% Drawing wires
% Line 1: 0 W
\draw[color=black] (0.000000,15.000000) -- (66.000000,15.000000);
% Line 2: 1 W
\draw[color=black] (0.000000,0.000000) -- (66.000000,0.000000);
% Done with wires; drawing gates
% Line 4: 0 G {$Z$}
\begin{scope}
\draw[fill=white] (12.000000, 15.000000) +(-45.000000:8.485281pt and 8.485281pt) -- +(45.000000:8.485281pt and 8.485281pt) -- +(135.000000:8.485281pt and 8.485281pt) -- +(225.000000:8.485281pt and 8.485281pt) -- cycle;
\clip (12.000000, 15.000000) +(-45.000000:8.485281pt and 8.485281pt) -- +(45.000000:8.485281pt and 8.485281pt) -- +(135.000000:8.485281pt and 8.485281pt) -- +(225.000000:8.485281pt and 8.485281pt) -- cycle;
\draw (12.000000, 15.000000) node {{$Z$}};
\end{scope}
% Line 5: 1 G {}
\begin{scope}
\draw[fill=white] (12.000000, -0.000000) +(-45.000000:8.485281pt and 8.485281pt) -- +(45.000000:8.485281pt and 8.485281pt) -- +(135.000000:8.485281pt and 8.485281pt) -- +(225.000000:8.485281pt and 8.485281pt) -- cycle;
\clip (12.000000, -0.000000) +(-45.000000:8.485281pt and 8.485281pt) -- +(45.000000:8.485281pt and 8.485281pt) -- +(135.000000:8.485281pt and 8.485281pt) -- +(225.000000:8.485281pt and 8.485281pt) -- cycle;
\draw (12.000000, -0.000000) node {{}};
\end{scope}
% Line 6: +1 0
\draw (33.000000,15.000000) -- (33.000000,0.000000);
\begin{scope}
\draw[fill=white] (33.000000, 0.000000) circle(3.000000pt);
\clip (33.000000, 0.000000) circle(3.000000pt);
\draw (30.000000, 0.000000) -- (36.000000, 0.000000);
\draw (33.000000, -3.000000) -- (33.000000, 3.000000);
\end{scope}
\filldraw (33.000000, 15.000000) circle(1.500000pt);
% Line 7: 0 G {$Z$}
\begin{scope}
\draw[fill=white] (54.000000, 15.000000) +(-45.000000:8.485281pt and 8.485281pt) -- +(45.000000:8.485281pt and 8.485281pt) -- +(135.000000:8.485281pt and 8.485281pt) -- +(225.000000:8.485281pt and 8.485281pt) -- cycle;
\clip (54.000000, 15.000000) +(-45.000000:8.485281pt and 8.485281pt) -- +(45.000000:8.485281pt and 8.485281pt) -- +(135.000000:8.485281pt and 8.485281pt) -- +(225.000000:8.485281pt and 8.485281pt) -- cycle;
\draw (54.000000, 15.000000) node {{$Z$}};
\end{scope}
% Line 8: 1 G {}
\begin{scope}
\draw[fill=white] (54.000000, -0.000000) +(-45.000000:8.485281pt and 8.485281pt) -- +(45.000000:8.485281pt and 8.485281pt) -- +(135.000000:8.485281pt and 8.485281pt) -- +(225.000000:8.485281pt and 8.485281pt) -- cycle;
\clip (54.000000, -0.000000) +(-45.000000:8.485281pt and 8.485281pt) -- +(45.000000:8.485281pt and 8.485281pt) -- +(135.000000:8.485281pt and 8.485281pt) -- +(225.000000:8.485281pt and 8.485281pt) -- cycle;
\draw (54.000000, -0.000000) node {{}};
\end{scope}
% Done with gates; drawing ending labels
% Done with ending labels; drawing cut lines and comments
% Done with comments
\end{tikzpicture}};
\node at (2.5, -1.8) {\begin{tikzpicture}[scale=1.000000,x=1pt,y=1pt]
\filldraw[color=white] (0.000000, -7.500000) rectangle (66.000000, 22.500000);
% Drawing wires
% Line 1: 0 W
\draw[color=black] (0.000000,15.000000) -- (66.000000,15.000000);
% Line 2: 1 W
\draw[color=black] (0.000000,0.000000) -- (66.000000,0.000000);
% Done with wires; drawing gates
% Line 4: 0 G {}
\begin{scope}
\draw[fill=white] (12.000000, 15.000000) +(-45.000000:8.485281pt and 8.485281pt) -- +(45.000000:8.485281pt and 8.485281pt) -- +(135.000000:8.485281pt and 8.485281pt) -- +(225.000000:8.485281pt and 8.485281pt) -- cycle;
\clip (12.000000, 15.000000) +(-45.000000:8.485281pt and 8.485281pt) -- +(45.000000:8.485281pt and 8.485281pt) -- +(135.000000:8.485281pt and 8.485281pt) -- +(225.000000:8.485281pt and 8.485281pt) -- cycle;
\draw (12.000000, 15.000000) node {{}};
\end{scope}
% Line 5: 1 G {$X$}
\begin{scope}
\draw[fill=white] (12.000000, -0.000000) +(-45.000000:8.485281pt and 8.485281pt) -- +(45.000000:8.485281pt and 8.485281pt) -- +(135.000000:8.485281pt and 8.485281pt) -- +(225.000000:8.485281pt and 8.485281pt) -- cycle;
\clip (12.000000, -0.000000) +(-45.000000:8.485281pt and 8.485281pt) -- +(45.000000:8.485281pt and 8.485281pt) -- +(135.000000:8.485281pt and 8.485281pt) -- +(225.000000:8.485281pt and 8.485281pt) -- cycle;
\draw (12.000000, -0.000000) node {{$X$}};
\end{scope}
% Line 6: +1 0
\draw (33.000000,15.000000) -- (33.000000,0.000000);
\begin{scope}
\draw[fill=white] (33.000000, 0.000000) circle(3.000000pt);
\clip (33.000000, 0.000000) circle(3.000000pt);
\draw (30.000000, 0.000000) -- (36.000000, 0.000000);
\draw (33.000000, -3.000000) -- (33.000000, 3.000000);
\end{scope}
\filldraw (33.000000, 15.000000) circle(1.500000pt);
% Line 7: 0 G {}
\begin{scope}
\draw[fill=white] (54.000000, 15.000000) +(-45.000000:8.485281pt and 8.485281pt) -- +(45.000000:8.485281pt and 8.485281pt) -- +(135.000000:8.485281pt and 8.485281pt) -- +(225.000000:8.485281pt and 8.485281pt) -- cycle;
\clip (54.000000, 15.000000) +(-45.000000:8.485281pt and 8.485281pt) -- +(45.000000:8.485281pt and 8.485281pt) -- +(135.000000:8.485281pt and 8.485281pt) -- +(225.000000:8.485281pt and 8.485281pt) -- cycle;
\draw (54.000000, 15.000000) node {{}};
\end{scope}
% Line 8: 1 G {$X$}
\begin{scope}
\draw[fill=white] (54.000000, -0.000000) +(-45.000000:8.485281pt and 8.485281pt) -- +(45.000000:8.485281pt and 8.485281pt) -- +(135.000000:8.485281pt and 8.485281pt) -- +(225.000000:8.485281pt and 8.485281pt) -- cycle;
\clip (54.000000, -0.000000) +(-45.000000:8.485281pt and 8.485281pt) -- +(45.000000:8.485281pt and 8.485281pt) -- +(135.000000:8.485281pt and 8.485281pt) -- +(225.000000:8.485281pt and 8.485281pt) -- cycle;
\draw (54.000000, -0.000000) node {{$X$}};
\end{scope}
% Done with gates; drawing ending labels
% Done with ending labels; drawing cut lines and comments
% Done with comments
\end{tikzpicture}};

\node at (5, 0.5) {\begin{tikzpicture}[scale=1.000000,x=1pt,y=1pt]
\filldraw[color=white] (0.000000, -7.500000) rectangle (48.000000, 7.500000);
% Drawing wires
% Line 1: 0 W
\draw[color=black] (0.000000,0.000000) -- (36.000000,0.000000);
\draw[color=black] (36.000000,-0.500000) -- (48.000000,-0.500000);
\draw[color=black] (36.000000,0.500000) -- (48.000000,0.500000);
% Done with wires; drawing gates
% Line 3: 0 G $Z$
\begin{scope}
\draw[fill=white] (12.000000, -0.000000) +(-45.000000:8.485281pt and 8.485281pt) -- +(45.000000:8.485281pt and 8.485281pt) -- +(135.000000:8.485281pt and 8.485281pt) -- +(225.000000:8.485281pt and 8.485281pt) -- cycle;
\clip (12.000000, -0.000000) +(-45.000000:8.485281pt and 8.485281pt) -- +(45.000000:8.485281pt and 8.485281pt) -- +(135.000000:8.485281pt and 8.485281pt) -- +(225.000000:8.485281pt and 8.485281pt) -- cycle;
\draw (12.000000, -0.000000) node {$Z$};
\end{scope}
% Line 4: 0 M {\scriptsize{$Z$}}
\draw[fill=white] (30.000000, -4.000000) -- (38.000000,-4.000000) arc (-90:90:4.000000pt) -- (30.000000,4.000000) -- cycle;
\draw (36.000000, 0.000000) node {{\scriptsize{$Z$}}};
% Done with gates; drawing ending labels
% Done with ending labels; drawing cut lines and comments
% Done with comments
\end{tikzpicture}};
\node at (5, -0.5) {\begin{tikzpicture}[scale=1.000000,x=1pt,y=1pt]
\filldraw[color=white] (0.000000, -7.500000) rectangle (48.000000, 7.500000);
% Drawing wires
% Line 1: 0 W
\draw[color=black] (0.000000,0.000000) -- (36.000000,0.000000);
\draw[color=black] (36.000000,-0.500000) -- (48.000000,-0.500000);
\draw[color=black] (36.000000,0.500000) -- (48.000000,0.500000);
% Done with wires; drawing gates
% Line 3: 0 G $X$
\begin{scope}
\draw[fill=white] (12.000000, -0.000000) +(-45.000000:8.485281pt and 8.485281pt) -- +(45.000000:8.485281pt and 8.485281pt) -- +(135.000000:8.485281pt and 8.485281pt) -- +(225.000000:8.485281pt and 8.485281pt) -- cycle;
\clip (12.000000, -0.000000) +(-45.000000:8.485281pt and 8.485281pt) -- +(45.000000:8.485281pt and 8.485281pt) -- +(135.000000:8.485281pt and 8.485281pt) -- +(225.000000:8.485281pt and 8.485281pt) -- cycle;
\draw (12.000000, -0.000000) node {$X$};
\end{scope}
% Line 4: 0 M {\scriptsize{$X$}}
\draw[fill=white] (30.000000, -4.000000) -- (38.000000,-4.000000) arc (-90:90:4.000000pt) -- (30.000000,4.000000) -- cycle;
\draw (36.000000, 0.000000) node {{\scriptsize{$X$}}};
% Done with gates; drawing ending labels
% Done with ending labels; drawing cut lines and comments
% Done with comments
\end{tikzpicture}};
\end{tikzpicture}
\caption{Elements of a generating set for the gauge group of a circuit comprised of \cnot{s}, $\ket{0}$/$\ket{+}$ preparations, and $Z$/$X$-basis measurements.}
\label{fig:OperatorBasis}
\end{figure}
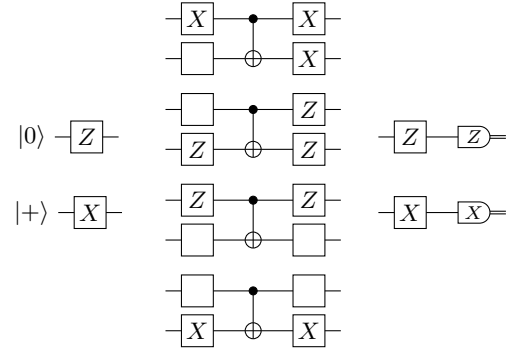

In addition to describing the times and places at which faults occur, these locations also support \emph{circuit gauge operators}, Pauli operators in the spacetime of the circuit which are equivalent to the identity (i.e. insertion of a circuit gauge operator into a circuit diagram preserves the action of the circuit on all possible input states). 
These operators form a group, which is often generated from a basis of low-weight operators that are local to each operation (preparation, gate, or measurement).
For completeness, we present this basis in \Cref{fig:OperatorBasis}.

A large subset of flag circuits (see, for example \cite{ReichardtFlags}) can be understood to measure these circuit gauge operators, in a sense which we now make precise.
We decompose a typical circuit with a flag measurement as follows:
\begin{figure}[!htbp]
\centering
\begin{tikzpicture}[scale=1.000000,x=1pt,y=1pt]
\filldraw[color=white] (0.000000, -7.500000) rectangle (210.000000, 52.500000);
% Drawing wires
% Line 1: 0 W \ket{+}
\draw[color=black] (0.000000,45.000000) -- (189.000000,45.000000);
\draw[color=black] (189.000000,44.500000) -- (210.000000,44.500000);
\draw[color=black] (189.000000,45.500000) -- (210.000000,45.500000);
\draw[color=black] (0.000000,45.000000) node[left] {$\ket{+}$};
% Line 2: 1 W
\draw[color=black] (0.000000,30.000000) -- (84.000000,30.000000);
\draw[color=black,dotted] (84.000000,30.000000) -- (114.000000,30.000000);
\draw[color=black,thin] (114.000000,30.000000) -- (181.500000,30.000000);
% Line 3: ...2 W
\draw[color=black] (0.000000,15.000000) node[anchor=mid east] {$\vdots$};
% Line 4: 3 W
\draw[color=black] (0.000000,0.000000) -- (84.000000,0.000000);
\draw[color=black,dotted] (84.000000,0.000000) -- (114.000000,0.000000);
\draw[color=black,thin] (114.000000,0.000000) -- (181.500000,0.000000);
% Done with wires; drawing gates
% Line 6: 1 3 G {$C_0$}
\draw (12.000000,30.000000) -- (12.000000,0.000000);
\begin{scope}
\draw[fill=white] (12.000000, 15.000000) +(-45.000000:8.485281pt and 29.698485pt) -- +(45.000000:8.485281pt and 29.698485pt) -- +(135.000000:8.485281pt and 29.698485pt) -- +(225.000000:8.485281pt and 29.698485pt) -- cycle;
\clip (12.000000, 15.000000) +(-45.000000:8.485281pt and 29.698485pt) -- +(45.000000:8.485281pt and 29.698485pt) -- +(135.000000:8.485281pt and 29.698485pt) -- +(225.000000:8.485281pt and 29.698485pt) -- cycle;
\draw (12.000000, 15.000000) node {{$C_0$}};
\end{scope}
% Line 7: 1 3 G {$P_0$} 0
\draw (36.000000,45.000000) -- (36.000000,0.000000);
\begin{scope}
\draw[fill=white] (36.000000, 15.000000) +(-45.000000:8.485281pt and 29.698485pt) -- +(45.000000:8.485281pt and 29.698485pt) -- +(135.000000:8.485281pt and 29.698485pt) -- +(225.000000:8.485281pt and 29.698485pt) -- cycle;
\clip (36.000000, 15.000000) +(-45.000000:8.485281pt and 29.698485pt) -- +(45.000000:8.485281pt and 29.698485pt) -- +(135.000000:8.485281pt and 29.698485pt) -- +(225.000000:8.485281pt and 29.698485pt) -- cycle;
\draw (36.000000, 15.000000) node {{$P_0$}};
\end{scope}
\filldraw (36.000000, 45.000000) circle(1.500000pt);
% Line 8: 1 3 G {$C_1$}
\draw (60.000000,30.000000) -- (60.000000,0.000000);
\begin{scope}
\draw[fill=white] (60.000000, 15.000000) +(-45.000000:8.485281pt and 29.698485pt) -- +(45.000000:8.485281pt and 29.698485pt) -- +(135.000000:8.485281pt and 29.698485pt) -- +(225.000000:8.485281pt and 29.698485pt) -- cycle;
\clip (60.000000, 15.000000) +(-45.000000:8.485281pt and 29.698485pt) -- +(45.000000:8.485281pt and 29.698485pt) -- +(135.000000:8.485281pt and 29.698485pt) -- +(225.000000:8.485281pt and 29.698485pt) -- cycle;
\draw (60.000000, 15.000000) node {{$C_1$}};
\end{scope}
% Line 9: 1:style=dotted 3:style=dotted G {$P_1$} 0
\draw (84.000000,45.000000) -- (84.000000,0.000000);
\begin{scope}
\draw[fill=white] (84.000000, 15.000000) +(-45.000000:8.485281pt and 29.698485pt) -- +(45.000000:8.485281pt and 29.698485pt) -- +(135.000000:8.485281pt and 29.698485pt) -- +(225.000000:8.485281pt and 29.698485pt) -- cycle;
\clip (84.000000, 15.000000) +(-45.000000:8.485281pt and 29.698485pt) -- +(45.000000:8.485281pt and 29.698485pt) -- +(135.000000:8.485281pt and 29.698485pt) -- +(225.000000:8.485281pt and 29.698485pt) -- cycle;
\draw (84.000000, 15.000000) node {{$P_1$}};
\end{scope}
\filldraw (84.000000, 45.000000) circle(1.500000pt);
% Line 11: 1:style=thin 3:style=thin G {$C_{s - 1}$} width=24
\draw (114.000000,30.000000) -- (114.000000,0.000000);
\begin{scope}
\draw[fill=white] (114.000000, 15.000000) +(-45.000000:16.970563pt and 29.698485pt) -- +(45.000000:16.970563pt and 29.698485pt) -- +(135.000000:16.970563pt and 29.698485pt) -- +(225.000000:16.970563pt and 29.698485pt) -- cycle;
\clip (114.000000, 15.000000) +(-45.000000:16.970563pt and 29.698485pt) -- +(45.000000:16.970563pt and 29.698485pt) -- +(135.000000:16.970563pt and 29.698485pt) -- +(225.000000:16.970563pt and 29.698485pt) -- cycle;
\draw (114.000000, 15.000000) node {{$C_{s - 1}$}};
\end{scope}
% Line 12: 1 3 G {$P_{s - 1}$} 0 width=24
\draw (150.000000,45.000000) -- (150.000000,0.000000);
\begin{scope}
\draw[fill=white] (150.000000, 15.000000) +(-45.000000:16.970563pt and 29.698485pt) -- +(45.000000:16.970563pt and 29.698485pt) -- +(135.000000:16.970563pt and 29.698485pt) -- +(225.000000:16.970563pt and 29.698485pt) -- cycle;
\clip (150.000000, 15.000000) +(-45.000000:16.970563pt and 29.698485pt) -- +(45.000000:16.970563pt and 29.698485pt) -- +(135.000000:16.970563pt and 29.698485pt) -- +(225.000000:16.970563pt and 29.698485pt) -- cycle;
\draw (150.000000, 15.000000) node {{$P_{s - 1}$}};
\end{scope}
\filldraw (150.000000, 45.000000) circle(1.500000pt);
% Line 13: 1 3 END
% Line 14: ...2 END
\draw[color=black] (189.000000,15.000000) node[fill=white,right,minimum height=15.000000pt,minimum width=15.000000pt,anchor = base,inner sep=0pt] {\phantom{$\;\vdots\;$}};
\draw[color=black] (189.000000,15.000000) node[anchor=mid west] {$\vdots$};
% Line 15: 0 M {\scriptsize $X$}
\draw[fill=white] (183.000000, 41.000000) -- (191.000000,41.000000) arc (-90:90:4.000000pt) -- (183.000000,49.000000) -- cycle;
\draw (189.000000, 45.000000) node {{\scriptsize $X$}};
% Done with gates; drawing ending labels
% Done with ending labels; drawing cut lines and comments
% Done with comments
\end{tikzpicture}
\caption{A circuit incorporating a flag measurement, composed of $s$ Clifford stages $\set{C_0 ,\, C_1 ,\, \ldots ,\, C_{s-1}}$, interleaved with controlled-Pauli gates $\set{P_0 ,\, P_1 ,\, \ldots ,\, P_{s-1}}$, with the control qubit initially prepared in the $\ket{+}$ state, and finally measured in the $X$ basis.}
\end{figure}
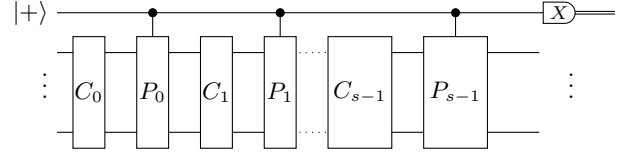
\FloatBarrier
A set of faults will result in a Pauli operator $E_j$ at the output of each Clifford stage $C_j$, incorporating both errors propagating from earlier stages, and the effects of additional faults within $C_j$.
Each $E_j$ propagates through the subsequent controlled-$P_j$ gate, resulting in a local $Z$ operator on the control qubit iff $\acom{E_j}{P_j}=0$. 
In order for the set of faults to be detected by the flag measurement, the number of these $Z$ operators must be odd. 
Using the well-known functional definition of commutation,
\begin{equation}
c(P, \, Q) = \begin{cases}
1 & \com{P}{Q} = 0 \\
-1 & \acom{P}{Q} = 0 
\end{cases}
\end{equation}
we can express this requirement formally.
The set of faults causing the operators $\set{E_j \vert j \in 0 \ldots s-1}$ will be detected by the flag circuit with controlled Pauli gates $\set{P_j \vert j \in 0 \ldots s-1}$ if
\begin{equation}
\prod_{j=0}^{s-1} c(E_j,\, P_j) = -1.
\end{equation}
Using the fact that, for Pauli operators $P$, $Q$, $R$, and $S$, $c(P \otimes Q ,\, R \otimes S) = c(P,\, R)c(Q,\, S)$, this criterion is equivalent to 
\begin{equation}
c \left(\bigotimes_{j=0}^{s-1} E_j ,\, \bigotimes_{j=0}^{s-1}P_j \right) = -1.
\end{equation}
Each of the Pauli operators $\bigotimes_{j=0}^{s-1} E_j$ and $\bigotimes_{j=0}^{s-1} P_j$ can be embedded within the set of locations of the circuit consisting only of the original Clifford stages $\set{C_j \vert j \in 0 \ldots s-1}$.  
For any decomposition of a given Clifford circuit into such stages, the Pauli $\bigotimes_{j=0}^{s-1} E_j$ can be calculated by restricting the \emph{spackle} of the fault set (defined in \cite{CircuitGaugeOperators}) to a subset of locations immediately after each stage $C_j$. 
This implies that the set of faults which are detected by a flag subcircuit (equivalent to a gauge operator measurement) is those whose spackles anticommute with the gauge operator from which the operators $P_j$ are derived. 

This is the main observation required to automate the construction of fault-tolerant circuits using flag measurements. 
In subsequent sections, we will make use of it in expressing the promotion problem as an integer linear program. 
However, this would become impractical for relatively small circuits if we were to attempt to optimize over all subsets of the entire gauge group of an input circuit. 
For this reason, we restrict our search to a subset of the gauge group, using a heuristic algorithm detailed in the following section.

\subsection{Gauge Operator Generation}

\begin{figure}[!htbp]
\centering
\begin{tikzpicture}
\node at (0, 0) {\begin{tikzpicture}[scale=1.000000,x=1pt,y=1pt]
\filldraw[color=white] (0.000000, -7.500000) rectangle (66.000000, 22.500000);
% Drawing wires
% Line 1: 0 W
\draw[color=black] (0.000000,15.000000) -- (66.000000,15.000000);
% Line 2: 1 W
\draw[color=black] (0.000000,0.000000) -- (66.000000,0.000000);
% Done with wires; drawing gates
% Line 4: 0 G {}
\begin{scope}
\draw[fill=white] (12.000000, 15.000000) +(-45.000000:8.485281pt and 8.485281pt) -- +(45.000000:8.485281pt and 8.485281pt) -- +(135.000000:8.485281pt and 8.485281pt) -- +(225.000000:8.485281pt and 8.485281pt) -- cycle;
\clip (12.000000, 15.000000) +(-45.000000:8.485281pt and 8.485281pt) -- +(45.000000:8.485281pt and 8.485281pt) -- +(135.000000:8.485281pt and 8.485281pt) -- +(225.000000:8.485281pt and 8.485281pt) -- cycle;
\draw (12.000000, 15.000000) node {{}};
\end{scope}
% Line 5: 1 G {$Z$}
\begin{scope}
\draw[fill=white] (12.000000, -0.000000) +(-45.000000:8.485281pt and 8.485281pt) -- +(45.000000:8.485281pt and 8.485281pt) -- +(135.000000:8.485281pt and 8.485281pt) -- +(225.000000:8.485281pt and 8.485281pt) -- cycle;
\clip (12.000000, -0.000000) +(-45.000000:8.485281pt and 8.485281pt) -- +(45.000000:8.485281pt and 8.485281pt) -- +(135.000000:8.485281pt and 8.485281pt) -- +(225.000000:8.485281pt and 8.485281pt) -- cycle;
\draw (12.000000, -0.000000) node {{$Z$}};
\end{scope}
% Line 6: +1 0
\draw (33.000000,15.000000) -- (33.000000,0.000000);
\begin{scope}
\draw[fill=white] (33.000000, 0.000000) circle(3.000000pt);
\clip (33.000000, 0.000000) circle(3.000000pt);
\draw (30.000000, 0.000000) -- (36.000000, 0.000000);
\draw (33.000000, -3.000000) -- (33.000000, 3.000000);
\end{scope}
\filldraw (33.000000, 15.000000) circle(1.500000pt);
% Line 7: 0 G {$Z$}
\begin{scope}
\draw[fill=white] (54.000000, 15.000000) +(-45.000000:8.485281pt and 8.485281pt) -- +(45.000000:8.485281pt and 8.485281pt) -- +(135.000000:8.485281pt and 8.485281pt) -- +(225.000000:8.485281pt and 8.485281pt) -- cycle;
\clip (54.000000, 15.000000) +(-45.000000:8.485281pt and 8.485281pt) -- +(45.000000:8.485281pt and 8.485281pt) -- +(135.000000:8.485281pt and 8.485281pt) -- +(225.000000:8.485281pt and 8.485281pt) -- cycle;
\draw (54.000000, 15.000000) node {{$Z$}};
\end{scope}
% Line 8: 1 G {$Z$}
\begin{scope}
\draw[fill=white] (54.000000, -0.000000) +(-45.000000:8.485281pt and 8.485281pt) -- +(45.000000:8.485281pt and 8.485281pt) -- +(135.000000:8.485281pt and 8.485281pt) -- +(225.000000:8.485281pt and 8.485281pt) -- cycle;
\clip (54.000000, -0.000000) +(-45.000000:8.485281pt and 8.485281pt) -- +(45.000000:8.485281pt and 8.485281pt) -- +(135.000000:8.485281pt and 8.485281pt) -- +(225.000000:8.485281pt and 8.485281pt) -- cycle;
\draw (54.000000, -0.000000) node {{$Z$}};
\end{scope}
% Done with gates; drawing ending labels
% Done with ending labels; drawing cut lines and comments
% Done with comments
\end{tikzpicture}};
\node at (1.75, 0) {$\mapsto$};
\node at (4, 0) {\begin{tikzpicture}[scale=1.000000,x=1pt,y=1pt]
\filldraw[color=white] (0.000000, -7.500000) rectangle (96.000000, 37.500000);
% Drawing wires
% Line 1: 0 W
\draw[color=black] (0.000000,30.000000) -- (96.000000,30.000000);
% Line 2: 1 W
\draw[color=black] (0.000000,15.000000) -- (96.000000,15.000000);
% Line 3: A W \ket{+}
\draw[color=black] (0.000000,0.000000) -- (84.000000,0.000000);
\draw[color=black] (84.000000,-0.500000) -- (96.000000,-0.500000);
\draw[color=black] (84.000000,0.500000) -- (96.000000,0.500000);
\draw[color=black] (0.000000,0.000000) node[left] {$\ket{+}$};
% Done with wires; drawing gates
% Line 5: 1 A
\draw (9.000000,15.000000) -- (9.000000,0.000000);
\filldraw (9.000000, 15.000000) circle(1.500000pt);
\filldraw (9.000000, 0.000000) circle(1.500000pt);
% Line 6: +1 0
\draw (27.000000,30.000000) -- (27.000000,15.000000);
\begin{scope}
\draw[fill=white] (27.000000, 15.000000) circle(3.000000pt);
\clip (27.000000, 15.000000) circle(3.000000pt);
\draw (24.000000, 15.000000) -- (30.000000, 15.000000);
\draw (27.000000, 12.000000) -- (27.000000, 18.000000);
\end{scope}
\filldraw (27.000000, 30.000000) circle(1.500000pt);
% Line 7: 0 A
\draw (45.000000,30.000000) -- (45.000000,0.000000);
\filldraw (45.000000, 30.000000) circle(1.500000pt);
\filldraw (45.000000, 0.000000) circle(1.500000pt);
% Line 8: 1 A
\draw (63.000000,15.000000) -- (63.000000,0.000000);
\filldraw (63.000000, 15.000000) circle(1.500000pt);
\filldraw (63.000000, 0.000000) circle(1.500000pt);
% Line 9: A M {\scriptsize{$X$}}
\draw[fill=white] (78.000000, -4.000000) -- (86.000000,-4.000000) arc (-90:90:4.000000pt) -- (78.000000,4.000000) -- cycle;
\draw (84.000000, 0.000000) node {{\scriptsize{$X$}}};
% Done with gates; drawing ending labels
% Done with ending labels; drawing cut lines and comments
% Done with comments
\end{tikzpicture}};
\end{tikzpicture}
\caption{Example of gauge-operator-to-flag-circuit mapping.}
\label{fig:gauge_operator_to_flag}
\end{figure}
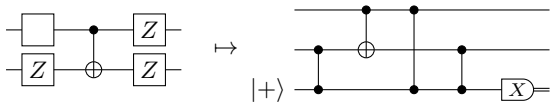

In order to transform a gauge operator into a measurement, we prepare an ancilla in the state $\ket{+}$, add a controlled-Pauli gate with the ancilla as the control and a target at every location where the gauge operator is supported, and conclude by measuring the ancilla in the $X$ basis, see \Cref{fig:gauge_operator_to_flag}.
\FloatBarrier
This implies that the number of high-error-rate operations (which we take to be two-qubit gates and measurements) required to measure the gauge operator in question is $w + 1$, where $w$ is the \emph{weight} of the gauge operator; the number of locations where it is not equal to the identity. 
For this reason, we only consider low-weight gauge operators. 
We generate these operators by iteratively expanding neighbourhoods in a graph calculated from the gauge group generators, see \Cref{alg:blob_generation}\footnote{subroutines generatorGraph and expandBlobs are described in \cref{app:subroutines}}.

\SetKwProg{Fn}{function}{}{end}
\SetKwFunction{flag_ops}{flag_ops}
\begin{algorithm}
\caption{Generation of low-weight gauge operators}
\label{alg:blob_generation}
\SetKwData{gens}{gens}
\SetKwData{maxWeight}{maxWeight}
\SetKwData{products}{products}
\SetKwData{genGraph}{genGraph}
\SetKwData{newProducts}{newProducts}
\SetKwData{findingNewProducts}{findingNewProducts}
\SetKwData{blobs}{blobs}
\SetKwData{blob}{blob}

\SetKwInOut{Input}{input}
\SetKwInOut{Output}{output}

\Input{\gens, \maxWeight}
\Output{\products}
\BlankLine

\blobs $=$ $\left[ \left[ q \right] \textbf{ for } q \textrm{ in } \mathds{Z}_{\abs{\textrm{\gens}}} \right]$\;
\genGraph $=$ generatorGraph(\gens)\;
\findingNewProducts $=$ true\;
\products $=$ \gens\;

\While{\findingNewProducts}{
\blobs $=$ expandBlobs(\blobs, \genGraph, \gens)\;
\newProducts $=$ filter($p \mapsto \textrm{weight}(p) \leq \textrm{\maxWeight}$, map(b $\rightarrow$ reduce($\times$, gens[b]), blobs))\;
\blobs $=$ filter($\textrm{\blob} \mapsto \textrm{weight}(\textrm{product}(\textrm{\gens}[\textrm{\blob}]) \leq \textrm{\maxWeight})$, \blobs)\;
\products = append(\products, \newProducts)\;
\findingNewProducts = not(isempty(\newProducts));
}
\products = filter($p \mapsto 1 < \textrm{weight}(p) < $ \maxWeight, \products)
\end{algorithm}
\FloatBarrier
The set of circuit gauge operators returned by this algorithm, while it is not guaranteed to be as large as possible, does contain the weight-two gauge operators corresponding to the low-overhead flag circuits in \cite{ReichardtFlags}. 
\section{Fault-Tolerant State Preparation by Integer Linear Programming}
\label{sec:MapToIntegerLinearProgram}

We have observed that flag circuits measure circuit gauge operators, detecting error sets whose spackles anticommute with those operators.
This implies that, in order to promote a circuit, we must find a set of circuit gauge operators $G$, each $g \in G$ anticommuting with a set of spackles $S_g$, such that $\bigcup_{g \in G}S_g \supseteq S$, where $S$ is the set of spackles that must be detected. 
This is a common problem in operations research, called \emph{set cover}. 
In the remainder of this work, we will study a related problem, \emph{weighted set cover}, in which each $g$ is associated with a non-negative cost, which we take to be $\textrm{weight}(g) + 1$, since each gauge operator will result in $\textrm{weight}(g)$ two-qubit gates and one measurement being added to the circuit in question\footnote{In the hardware we intend to use, SPAM and two-qubit operations have similar error rates. A cost function of the form $A \cdot \textrm{weight}(g) + B$ would be more appropriate for dissimilar rates.}.
We solve the resulting weighted set cover problems by expressing them as integer linear programs of the form:
\begin{flalign}
\argmin_{\vec{g}} \vec{c} \cdot \vec{g} \nonumber \\
\textrm{such that } & \vec{s} \cdot \vec g > 0 \,\, \forall \,\, \vec{s} \in S,
\end{flalign}
where $\vec{g} \in \mathds{Z}_2^{\abs{G}}$ is 1 for each $g$ being measured, $\vec{c}_g = \textrm{weight}(g) + 1$, and $S$ contains one $\vec{s}$ for each spackle $s$ which must be detected. 
In order to determine the set of spackles which must be detected, we spackle all faults that occur with probability $O(p)$, calculate the set of products of size $t$ (the number of faults to be detected), and remove those that are detected by existing measurements or are equivalent to an operator of weight $\leq t$ at the output, up to multiplication by a stabilizer of the state being prepared. 

Once such a set of operators is found, it is used to augment the input circuit as in \cref{fig:gauge_operator_to_flag}. 
This ensures that combinations of faults in the input circuit will be detected if they would otherwise propagate to an undetected high-weight error at the output.
Attempts at state preparation in which such errors are detected can then be post-selected.
However, this is not sufficient to produce states fault-tolerantly, because faults in the flag sub-circuit being constructed are not accounted for during optimization.

\begin{figure}[!htbp]
\centering
\begin{tikzpicture}[scale=1.000000,x=1pt,y=1pt]
\filldraw[color=white] (0.000000, -7.500000) rectangle (144.000000, 52.500000);
% Drawing wires
% Line 1: anc W \ket{+}
\draw[color=black] (0.000000,45.000000) -- (132.000000,45.000000);
\draw[color=black] (132.000000,44.500000) -- (144.000000,44.500000);
\draw[color=black] (132.000000,45.500000) -- (144.000000,45.500000);
\draw[color=black] (0.000000,45.000000) node[left] {$\ket{+}$};
% Line 2: 0 W
\draw[color=black] (0.000000,30.000000) -- (144.000000,30.000000);
% Line 3: 1 W
\draw[color=black] (0.000000,15.000000) -- (144.000000,15.000000);
% Line 4: 2 W
\draw[color=black] (0.000000,0.000000) -- (144.000000,0.000000);
% Done with wires; drawing gates
% Line 6: 0 G $0$
\begin{scope}
\draw[fill=white] (12.000000, 30.000000) +(-45.000000:8.485281pt and 8.485281pt) -- +(45.000000:8.485281pt and 8.485281pt) -- +(135.000000:8.485281pt and 8.485281pt) -- +(225.000000:8.485281pt and 8.485281pt) -- cycle;
\clip (12.000000, 30.000000) +(-45.000000:8.485281pt and 8.485281pt) -- +(45.000000:8.485281pt and 8.485281pt) -- +(135.000000:8.485281pt and 8.485281pt) -- +(225.000000:8.485281pt and 8.485281pt) -- cycle;
\draw (12.000000, 30.000000) node {$0$};
\end{scope}
% Line 7: 1 G $1$
\begin{scope}
\draw[fill=white] (12.000000, 15.000000) +(-45.000000:8.485281pt and 8.485281pt) -- +(45.000000:8.485281pt and 8.485281pt) -- +(135.000000:8.485281pt and 8.485281pt) -- +(225.000000:8.485281pt and 8.485281pt) -- cycle;
\clip (12.000000, 15.000000) +(-45.000000:8.485281pt and 8.485281pt) -- +(45.000000:8.485281pt and 8.485281pt) -- +(135.000000:8.485281pt and 8.485281pt) -- +(225.000000:8.485281pt and 8.485281pt) -- cycle;
\draw (12.000000, 15.000000) node {$1$};
\end{scope}
% Line 8: 2 G $2$
\begin{scope}
\draw[fill=white] (12.000000, -0.000000) +(-45.000000:8.485281pt and 8.485281pt) -- +(45.000000:8.485281pt and 8.485281pt) -- +(135.000000:8.485281pt and 8.485281pt) -- +(225.000000:8.485281pt and 8.485281pt) -- cycle;
\clip (12.000000, -0.000000) +(-45.000000:8.485281pt and 8.485281pt) -- +(45.000000:8.485281pt and 8.485281pt) -- +(135.000000:8.485281pt and 8.485281pt) -- +(225.000000:8.485281pt and 8.485281pt) -- cycle;
\draw (12.000000, -0.000000) node {$2$};
\end{scope}
% Line 10: 0 G $P$ anc % \textit{i}
\draw (36.000000, 52.500000) node[text width=144pt,above,text centered] {\textit{i}};
\draw (36.000000,45.000000) -- (36.000000,30.000000);
\begin{scope}
\draw[fill=white] (36.000000, 30.000000) +(-45.000000:8.485281pt and 8.485281pt) -- +(45.000000:8.485281pt and 8.485281pt) -- +(135.000000:8.485281pt and 8.485281pt) -- +(225.000000:8.485281pt and 8.485281pt) -- cycle;
\clip (36.000000, 30.000000) +(-45.000000:8.485281pt and 8.485281pt) -- +(45.000000:8.485281pt and 8.485281pt) -- +(135.000000:8.485281pt and 8.485281pt) -- +(225.000000:8.485281pt and 8.485281pt) -- cycle;
\draw (36.000000, 30.000000) node {$P$};
\end{scope}
\filldraw (36.000000, 45.000000) circle(1.500000pt);
% Line 11: 1 G $Q$ anc % \textit{ii}
\draw (60.000000, 52.500000) node[text width=144pt,above,text centered] {\textit{ii}};
\draw (60.000000,45.000000) -- (60.000000,15.000000);
\begin{scope}
\draw[fill=white] (60.000000, 15.000000) +(-45.000000:8.485281pt and 8.485281pt) -- +(45.000000:8.485281pt and 8.485281pt) -- +(135.000000:8.485281pt and 8.485281pt) -- +(225.000000:8.485281pt and 8.485281pt) -- cycle;
\clip (60.000000, 15.000000) +(-45.000000:8.485281pt and 8.485281pt) -- +(45.000000:8.485281pt and 8.485281pt) -- +(135.000000:8.485281pt and 8.485281pt) -- +(225.000000:8.485281pt and 8.485281pt) -- cycle;
\draw (60.000000, 15.000000) node {$Q$};
\end{scope}
\filldraw (60.000000, 45.000000) circle(1.500000pt);
% Line 12: 0 1 2 G $C$
\draw (84.000000,30.000000) -- (84.000000,0.000000);
\begin{scope}
\draw[fill=white] (84.000000, 15.000000) +(-45.000000:8.485281pt and 29.698485pt) -- +(45.000000:8.485281pt and 29.698485pt) -- +(135.000000:8.485281pt and 29.698485pt) -- +(225.000000:8.485281pt and 29.698485pt) -- cycle;
\clip (84.000000, 15.000000) +(-45.000000:8.485281pt and 29.698485pt) -- +(45.000000:8.485281pt and 29.698485pt) -- +(135.000000:8.485281pt and 29.698485pt) -- +(225.000000:8.485281pt and 29.698485pt) -- cycle;
\draw (84.000000, 15.000000) node {$C$};
\end{scope}
% Line 13: 2 G $R$ anc % \textit{iii}
\draw (108.000000, 52.500000) node[text width=144pt,above,text centered] {\textit{iii}};
\draw (108.000000,45.000000) -- (108.000000,0.000000);
\begin{scope}
\draw[fill=white] (108.000000, -0.000000) +(-45.000000:8.485281pt and 8.485281pt) -- +(45.000000:8.485281pt and 8.485281pt) -- +(135.000000:8.485281pt and 8.485281pt) -- +(225.000000:8.485281pt and 8.485281pt) -- cycle;
\clip (108.000000, -0.000000) +(-45.000000:8.485281pt and 8.485281pt) -- +(45.000000:8.485281pt and 8.485281pt) -- +(135.000000:8.485281pt and 8.485281pt) -- +(225.000000:8.485281pt and 8.485281pt) -- cycle;
\draw (108.000000, -0.000000) node {$R$};
\end{scope}
\filldraw (108.000000, 45.000000) circle(1.500000pt);
% Line 15: TOUCH
% Line 16: anc M {\scriptsize{$X$}}
\draw[fill=white] (126.000000, 41.000000) -- (134.000000,41.000000) arc (-90:90:4.000000pt) -- (126.000000,49.000000) -- cycle;
\draw (132.000000, 45.000000) node {{\scriptsize{$X$}}};
% Line 17: 0 G $3$
\begin{scope}
\draw[fill=white] (132.000000, 30.000000) +(-45.000000:8.485281pt and 8.485281pt) -- +(45.000000:8.485281pt and 8.485281pt) -- +(135.000000:8.485281pt and 8.485281pt) -- +(225.000000:8.485281pt and 8.485281pt) -- cycle;
\clip (132.000000, 30.000000) +(-45.000000:8.485281pt and 8.485281pt) -- +(45.000000:8.485281pt and 8.485281pt) -- +(135.000000:8.485281pt and 8.485281pt) -- +(225.000000:8.485281pt and 8.485281pt) -- cycle;
\draw (132.000000, 30.000000) node {$3$};
\end{scope}
% Line 18: 1 G $4$
\begin{scope}
\draw[fill=white] (132.000000, 15.000000) +(-45.000000:8.485281pt and 8.485281pt) -- +(45.000000:8.485281pt and 8.485281pt) -- +(135.000000:8.485281pt and 8.485281pt) -- +(225.000000:8.485281pt and 8.485281pt) -- cycle;
\clip (132.000000, 15.000000) +(-45.000000:8.485281pt and 8.485281pt) -- +(45.000000:8.485281pt and 8.485281pt) -- +(135.000000:8.485281pt and 8.485281pt) -- +(225.000000:8.485281pt and 8.485281pt) -- cycle;
\draw (132.000000, 15.000000) node {$4$};
\end{scope}
% Line 19: 2 G $5$
\begin{scope}
\draw[fill=white] (132.000000, -0.000000) +(-45.000000:8.485281pt and 8.485281pt) -- +(45.000000:8.485281pt and 8.485281pt) -- +(135.000000:8.485281pt and 8.485281pt) -- +(225.000000:8.485281pt and 8.485281pt) -- cycle;
\clip (132.000000, -0.000000) +(-45.000000:8.485281pt and 8.485281pt) -- +(45.000000:8.485281pt and 8.485281pt) -- +(135.000000:8.485281pt and 8.485281pt) -- +(225.000000:8.485281pt and 8.485281pt) -- cycle;
\draw (132.000000, -0.000000) node {$5$};
\end{scope}
% Done with gates; drawing ending labels
% Done with ending labels; drawing cut lines and comments
% Done with comments
\end{tikzpicture}
\caption{Bare-ancilla measurement of a weight-3 circuit gauge operator $P \otimes Q \otimes R$. 
To facilitate analysis, input and output locations are numbered explicitly.}
\label{fig:weight_3_flag}
\end{figure}
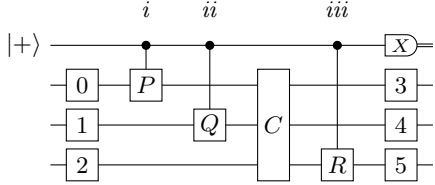

When promoting a circuit from tolerating zero faults to $t - 1$, this problem can be temporarily ignored until the final stage, in which it is beneficial to guarantee fault tolerance a priori. 
This can be accomplished by limiting the weight of circuit gauge operators considered to be at most three, and the \emph{output weight} (number of locations at the output of the circuit in the support of the gauge operator) to be at most one. 
To see this, we consider the generic circuit in \Cref{fig:weight_3_flag}.
\FloatBarrier
In order to prove that a measurement of this type does not produce high-weight faults not seen in the original circuit $C$, we first consider the effects of Pauli generators occurring after each of the gates.
Note that, in contrast to the usual method of analysis, where Paulis are always propagated to later times, we will analyse the circuit faults in \cref{fig:weight_3_flag} by finding equivalent Paulis at the input and output locations (presented in \Cref{tab:equivalent_faults}), propagating forwards or backwards in time where convenient.
\begin{table}[!htbp]
\centering
\begin{tabular}{c|ccc}
Gate & Generator & Equivalent Pauli & Syndrome \\
\hline
\textit{i} 		& $XI$ 			& $Q_1 R_5$ & 0 \\
				& $ZI$ 			& $\mathds{1}$ & 1\\
				& $IP$ 			& $P_0$ & 0\\
				& $IP^{\perp}$ 	& $P^{\perp}_0$ & 1 \\
\hline
\textit{ii}		& $XI$ 			& $R_5$ & 0 \\
				& $ZI$ 			& $\mathds{1}$ & 1\\
				& $IQ$ 			& $Q_1$ & 0\\
				& $IQ^{\perp}$ 	& $Q^{\perp}_1$ & 1\\
\hline
\textit{iii}	& $XI$ 			& $R_5$ & 0 \\
				& $ZI$ 			& $\mathds{1}$ & 1\\
				& $IR$ 			& $R_5$ & 0\\
				& $IR^{\perp}$ 	& $R^{\perp}_5$ & 0
\end{tabular}
\caption{Equivalent Pauli operators and syndromes for local generators occurring after each two-qubit gate in \cref{fig:weight_3_flag}. }
\label{tab:equivalent_faults}
\end{table}
\FloatBarrier
Since any product of local generators may occur after a two-qubit gate, there is a possibility that the operators $P_0 Q_1 R_5$ or $Q_1 R_5$ may be equivalent to an error occurring with probability $O(p)$. 
However, the gauge operator being measured is equivalent to $P_0 Q_1 R_5$ in the resulting circuit, which implies that $P_0 Q_1 R_5 \sim \mathds{1}$ and $Q_1 R_5 \sim P_0$.
Therefore, all first-order errors in the flag circuit are equivalent to first-order errors which could occur in the original circuit $C$, and fault tolerance of the output circuit is guaranteed, provided that the measured gauge operators anticommute with all necessary spackles.

\section{Examples}
\label{sec:Examples}
It is not clear a priori how best to use an algorithm for circuit promotion to create fault-tolerant state preparation circuits.
It may be preferable to promote incrementally, adding as few gates as possible in each stage, in order to reduce the number of unaccounted for faults to be detected in subsequent stages, or it may be preferable to first attempt to detect all dangerous sets of faults up to size $t$, potentially resulting in a smaller set of measurements, and fewer faults to be detected by weight-three flags. 
In addition, it may be preferable to include high-weight stabilizers (both at the output of the circuit, and between depth-one layers) in the set of measurable operators, in addition to the weight-restricted gauge operators generated in \cref{alg:blob_generation}.
In this section, we compare the performance of strategies incorporating incremental/one-step promotion and low/high-weight operators, for CSS codes with distances 3-7, at block sizes up to $n=42$.

Noting that integer linear programming is an NP-complete problem, we restrict our search for state preparation circuits to those which allow a solution in terms of a series of small integer linear programs. 
Firstly, we only consider CSS codes with non-fault-tolerant initial circuits expressed in terms of $\ket{0}$ preparation, $\ket{+}$ preparation, and CNot, separating the gauge group into operators consisting only of $X$/$I$ and those consisting only of $Z$/$I$. 
This allows us to alternate between promotion steps that increase the tolerance to $X$ faults and $Z$ faults independently, as in \cite{PehamSAT}, at the cost of solving each circuit promotion problem twice, once with promotion targeting $X$ faults first, and once with $Z$ faults.
Secondly, we begin our search with non-fault-tolerant circuits generated using beam search over graph states \cite{AI_Circuit_Generation} for codes which do not permit small preparation circuits to be constructed by hand, in order to limit the number of faults and locations under consideration. 
We compare these non-fault-tolerant and fault-tolerant circuits to the prior state of the art for reference.
Finally, for the larger codes under consideration, we only use search strategies which produce optimal circuits for smaller codes, as runtime and memory consumption prohibit the exploration of the full set of strategies.

Our results are summarized in \cref{tab:gate_counts}.
\onecolumngrid
\begin{center}
\begin{table}[!htbp]
\begin{tabular}{c|c|c|c|c|c|c|c}
Code & \makecell{Non-FT\\Circuit} & Weights & Strategy & \makecell{$n_{\textrm{ops}}$\\(non-FT)} & \makecell{SOTA\\(non-FT)} & \makecell{$n_{\textrm{ops}}$\\(FT)} & \makecell{SOTA\\(FT)}  \\
\hline
$\nkd{7}{1}{3}$ & BS & $(3,\,3,\,3)$ & one-step & 8 & 8 \cite{GotoSteane} \footnote{This optimization results in a single measurement of a weight-3 gauge operator, guaranteeing that other strategies would return the same result.}& 12 & 12 \cite{GotoSteane} \\
$\nkd{12}{2}{4}$ & BS & $(6,\,12,\,12)$ & promotion & 16 & 16 \cite{PehamSAT} & 43 & 43 \cite{CarbonCode} \\
$d=4$ surface code & by-hand & $(4,\,8,\,12)$ & one-step & 17 & 17 \cite{PehamSAT} & 38 & 52 \cite{FlagAtOrigin} \\
$\nkd{16}{2}{4}$ colour code \cite{PrabhuReichardt} & by-hand & $(4,\,8,\,12)$ & one-step & 21 & 21 \cite{PehamSAT} & 52 & 92 \cite{FlagAtOrigin} \\ 
$\nkd{16}{4}{4}$ colour code \cite{PrabhuReichardt} & by-hand & $(4,\,8,\,12)$ & promotion & 22 & 22 \cite{PehamSAT} & 56 & 96 \cite{FlagAtOrigin} \\ 
$\nkd{16}{4}{4}$ low-weight & BS & $(4,\,8,\,12)$ & promotion & 24 & 27 \cite{PehamSAT} & 59 & 96 \cite{FlagAtOrigin} \\ 
$\nkd{16}{6}{4}$ Reed-Muller & by-hand & $(4,\,8,\,12)$ & promotion & 25 & 25 & 63 & 116 \cite{FlagAtOrigin} \\ 
$\nkd{24}{10}{4}$ 2BGA code & BS & $(4,\,8,\,12)$ & promotion & 39 & 41 \cite{PehamSAT} & 89 & 170 \cite{FlagAtOrigin} \\ 
$\nkd{17}{1}{5}$ colour code & BS & $(6,\,17,\,17)$ & one-step & 23 & 23 \cite{PehamSAT} & 62 & 84 \cite{PehamSAT} \\ 
$\nkd{24}{4}{5}$ & BS & $(4,\,8,\,12)$ & promotion & 43 & 54 \cite{PehamSAT} & 147 & 205 \cite{FlagAtOrigin} \\ 
$\nkd{20}{2}{6}$ & BS & $(4,\,8,\,12)$ & promotion & 32 & 32 \cite{PehamSAT} & 166 & 192 \cite{FlagAtOrigin}\\ 
$\nkd{42}{10}{6}$ & BS & $(4,\,8,\,12)$ & promotion & 91 & 93 & 350 & 613 \cite{FlagAtOrigin} \\ 
$\nkd{23}{1}{7}$ & BS & $(4,\,8,\,12)$ & promotion & 45 & 54 \cite{SelwynCommunication} & 248 & 306 \cite{PehamSAT}  
\end{tabular}
\caption{Gate counts for ILP-derived circuits. 
ILP is able to meet or surpass the state of the art for all cases considered.
Reduction in gate count is greater for high-rate codes.}
\label{tab:gate_counts}
\end{table}
\end{center}
\twocolumngrid

\subsection{Simulation \& Hardware Results}
\label{sec:Results}
\begin{figure}[!htbp]
\centering
\includegraphics[width=0.8\columnwidth]{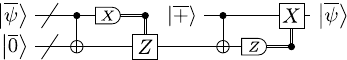}
\caption{Steane error correction cycle rendered robust to leakage using logical one-bit teleportation \cite{AliferisTerhal, CarbonCode}.}
\label{fig:leakage_robust_steane}
\end{figure}

In order to assess the utility of the state preparation circuits derived above, we incorporate them into Steane error correction, using logical one-way teleportation to ensure that leaked ions are regularly removed and replaced with ions recently prepared in computational states, see \cref{fig:leakage_robust_steane}.
To assess performance in a regime close to commercial applicability, we use the $\nkd{24}{10}{4}$ code, as it has a coding rate near $\nicefrac{1}{2}$, while allowing two blocks to be stored in Quantinuum's System Model H2 quantum computer, with sufficient free qubits to use as ancillae for state preparation, after post-processing for ancilla reuse \cite{AncillaReuse}. 
This requires the $\overline{\ket{+}^{10}}$ state to be prepared in addition to the $\overline{\ket{0}^{10}}$ state, see \cref{fig:24_10_4_prep} for both state preparations.

We prepare the logical zero state fault-tolerantly, perform quantum error correction, and measure destructively in the $Z$ basis, repeating for $10^4$ shots. 
A breakdown of the different results obtained is given in \cref{tab:exptl_results}.
\begin{center}
\begin{table}[!htbp]
\begin{tabular}{l|c|c|c|c|c|c}
& Pre-selected & $w_d=0$ & $w_d=1$ & $w_d=2$ & $w_d=3$ & $w_d=4$ \\
\hline
H2-2 & $2883_{44.6}^{45.5}$	& $5343_{49.7}^{49.5}$ & $1614_{35.9}^{37.3}$ & $159_{11.5}^{13.4}$ & $1_{0.288}^{2.28}$ & $0_{0}^{1.8}$ \\ 
H2-2E & $22593_{131}^{132}$	& $64347_{151}^{150}$ & $12407_{103}^{104}$ & $650_{24.3}^{26.3}$ & $3_{0.907}^{2.90}$ & $0_{0}^{1.8}$ \\
\end{tabular}
\caption{Results of $10^4$ $(10^5)$ independent prepare-QEC-and-measure experiments on H2-2 (H2-2E) using a one-way-teleportation-based variation of Steane QEC on the $\nkd{24}{10}{4}$ two-block group algebra code.
Shots are classified by the maximum error weight assigned by a lookup table decoder: $w_d = 0$ corresponds to no error detected, $w_d=1$ to a correctable error, $w_d=2$ to an uncorrectable error, $w_d=3$ to an error misidentified as correctable (identified as a logical bitflip at the end of the computation), and $w_d=4$ to an unintentional logical operator occurring without resulting in a syndrome. 
The presence of one logical error in 7117 accepted shots implies a logical error rate of $\nicefrac{1}{7117} \sim 1.4 \times 10^{-4}$ on the block, $\sim 1.4 \times 10^{-5}$ per logical qubit.
Results are statistically distinguishable, but qualitatively similar, and indicate that error rates on the device may be $\sim 25\%$ higher during the experiment, as compared to calibration.}
\label{tab:exptl_results}
\end{table}
\end{center}
\FloatBarrier

\section{Discussion \& Conclusion}
\label{sec:DiscussionAndConclusion}
This work demonstrates a practical use case of the circuit gauge operator formalism, the automation of the construction of flag fault-tolerant circuits through repeated application of combinatorial optimization. 
While the resulting circuits are smaller than the state-of-the-art alternatives in the literature, we suspect that much smaller circuits are possible, for a variety of reasons. 
Firstly, the separation of fault-tolerant circuit design into two stages (non-fault-tolerant circuit design and subsequent promotion) is not technically necessary, though it is not immediately clear how to unite these two stages into a single process that begins from an abstract state preparation task and results in a fault-tolerant circuit.
Secondly, once a set of gauge operators has been derived whose measurement would result in fault tolerance, bare-ancilla measurement is not necessarily optimal. 
The fact that each of these operators results in a non-fault-tolerant subcircuit which in turn requires further gauge operators to be measured implies that the cost function used to derive them is inaccurate, and this inaccuracy compounds for higher-distance codes. 
In addition, stabilizer measurement techniques that use independent sub-circuits for each operator (e.g. Shor) are known to be more expensive and less reliable than those that measure multiple operators simultaneously (e.g. Steane and Knill), provided that the device to be used has sufficient connectivity to be able to prepare stabilizer states to be consumed within the measurement. 
Incorporating techniques such as Huang \& Brown's unified measurement scheme \cite{BetweenShorAndSteane}, therefore, represents a promising avenue for future research.

Another promising avenue is the application of gauge operator measurement to logical operations, such as fold-transversal gates \cite{FoldTransversalGates}. 
For codes with distances greater than two, this requires the use or construction of a decoding algorithm as part of the exit criterion of the promotion algorithm, which is complex, but achievable in principle.
Similarly, it is possible in principle to promote non-fault-tolerant circuits for tasks other than state preparation, at the cost of incorporating more complicated fault tolerance criteria.

In addition to these factors, it is also desirable to track the circuit depth and number of ancillae used in the proposed circuit as it is being promoted, and to insert a small number of sub-circuits in each round, re-evaluating the sets of dangerous errors more frequently, in order to avoid adding large sets of high-weight errors simultaneously. 
These factors combined indicate that the promotion problem may be more productively phrased as a Markov decision process, at which point artificial-intelligence-based search algorithms may be used, as they are to derive non-fault-tolerant circuits 
\cite{AI_Circuit_Generation}. 

\begin{acknowledgments}
BC acknowledges useful discussions with Vadym Kliuchnikov, and Peter-Jan Derks.
BC thanks Matthew Girling, Matt DeCross and Eli Chertkov for assistance with compilation for ancilla reuse, dynamical decoupling, and circuit visualization.
This research was supported by the Army Research Office through W911NF253A0002.
\end{acknowledgments}

\clearpage
\bibliographystyle{apsrev4-1}
\bibliography{HighDistanceStatePrep}

\appendix

\section{Subroutines}
\label{app:subroutines}
\begin{algorithm}[!htbp]
\caption{generatorGraph: constructing a graph from gauge group generators, to be used to generate low-weight products}
\label{alg:generatorGraph}
\SetKwInOut{Input}{input}
\SetKwInOut{Output}{output}

\SetKwData{gens}{gens}
\SetKwData{nGens}{nGens}
\SetKwData{uIndex}{uIndex}
\SetKwData{vIndex}{vIndex}
\SetKwData{genGraph}{genGraph}
\SetKwData{support}{support}

\Input{\gens}
\Output{\genGraph}	
	
	\nGens $=$ length(\gens)\;
	\genGraph = newGraph(\nGens)\;
	\For{\uIndex $\in \left[ 1 \ldots \nGens - 1 \right]$}{
		\For{\vIndex $\in \left[ \uIndex + 1 \ldots \nGens \right]$}{
			\If{$\gens[\uIndex] \cap \gens[\vIndex] \neq \emptyset$}{
				addEdge(\genGraph, (\uIndex, \vIndex))\;
			}		
		}	
	}
\end{algorithm}
\FloatBarrier

\begin{algorithm}[!htbp]
\caption{expandBlobs: creating new neighbourhoods of the generator graph to derive new low-weight products of gauge generators.}
\label{alg:expandBlobs}

\SetKwInOut{Input}{input}
\SetKwInOut{Output}{output}

\SetKwData{gens}{gens}
\SetKwData{blob}{blob}
\SetKwData{blobs}{blobs}
\SetKwData{prodd}{prod}
\SetKwData{dx}{dx}
\SetKwData{newBlobs}{newBlobs}
\SetKwData{newestBlobs}{newerBlobs}
\SetKwData{genGraph}{genGraph}
\SetKwData{neighbours}{nbrs}
\SetKwData{neighbourhood}{neighbourhood}

\Input{\blobs, \genGraph, \gens}
\Output{\newBlobs}

\newBlobs = $\emptyset$\;
\For{\blob $\in$ \blobs}{
	\prodd = $\prod_{\dx \in \blob} \gens[\dx]$\;
	\neighbours = $\bigcup_{\dx \in \blob} \neighbourhood(\genGraph,\,\dx)$\;
	\neighbours = $\neighbours \setminus \blob$\;
	\tcc{Only vertices that actually overlap the product should be allowed into the new blob}
	\neighbours = filter($\dx \mapsto $overlapping$(\gens[\dx], \prodd)$, \neighbours)\;
	\newestBlobs = unique(map($\dx \mapsto \blob \cup \left \lbrace \dx \right \rbrace$, \neighbours))\;
	\newBlobs = \newBlobs $\cup$ \newestBlobs\;
}
\end{algorithm}

\section{Stabilizer Tableaux}
\subsection{$\nkd{16}{4}{4}$}
We obtain a $\nkd{16}{4}{4}$ code with weight-six stabilizers by projecting two of the logical qubits of the $\nkd{16}{6}{4}$ Reed-Muller code into a logical Bell pair:
\setcounter{MaxMatrixCols}{16}
\begin{flalign}
S_{\nkd{16}{4}{4}} &= \left \langle   X_{3} X_{4} X_{6} X_{8} X_{10} X_{11},\, X_{1} X_{2} X_{6} X_{8} X_{9} X_{12},\, \right. \nonumber \\
& \left. X_{6} X_{7} X_{10} X_{12} X_{15} X_{16},\, X_{6} X_{7} X_{9} X_{11} X_{13} X_{14},\, \right. \nonumber \\
& \left. X_{1} X_{3} X_{5} X_{6} X_{14} X_{15},\, X_{1} X_{2} X_{5} X_{7} X_{10} X_{11},\, \right. \nonumber \\
& \left. Z_{3} Z_{4} Z_{6} Z_{8} Z_{10} Z_{11},\, Z_{1} Z_{2} Z_{6} Z_{8} Z_{9} Z_{12},\, \right. \nonumber \\
& \left. Z_{6} Z_{7} Z_{10} Z_{12} Z_{15} Z_{16},\, Z_{6} Z_{7} Z_{9} Z_{11} Z_{13} Z_{14},\, \right. \nonumber \\
& \left. Z_{1} Z_{3} Z_{5} Z_{6} Z_{14} Z_{15},\, Z_{1} Z_{2} Z_{5} Z_{7} Z_{10} Z_{11} \right \rangle \label{eq:16_4_4_6_code}
\end{flalign}

\subsection{Two-Block Group Algebra Codes}
The $\nkd{24}{10}{4}$, $\nkd{24}{4}{5}$, and $\nkd{42}{10}{6}$ codes studied are two-block group algebra codes \cite{2BGA}, also available at \texttt{qecdb.org} \cite{qecdb}:
\begin{flalign}
S_{\nkd{24}{10}{4}} = & \left \langle X_{1} X_{3} X_{6} X_{8} X_{13} X_{17} X_{18} X_{23}, \right. \nonumber \\
& \left. X_{2} X_{4} X_{5} X_{11} X_{14} X_{18} X_{20} X_{23}, \right. \nonumber \\
& \left. X_{3} X_{7} X_{10} X_{11} X_{15} X_{18} X_{22} X_{24}, \right. \nonumber \\
& \left. X_{4} X_{6} X_{9} X_{12} X_{16} X_{22} X_{23} X_{24}, \right. \nonumber \\
& \left. X_{2} X_{3} X_{5} X_{6} X_{13} X_{15} X_{16} X_{17}, \right. \nonumber \\
& \left. X_{4} X_{6} X_{7} X_{10} X_{15} X_{18} X_{19} X_{21}, \right. \nonumber \\
& \left. X_{1} X_{5} X_{7} X_{12} X_{13} X_{14} X_{19} X_{22}, \right. \nonumber \\
& \left. Z_{1} Z_{5} Z_{7} Z_{9} Z_{13} Z_{19} Z_{20} Z_{22}, \right. \nonumber \\
& \left. Z_{2} Z_{7} Z_{8} Z_{9} Z_{14} Z_{17} Z_{21} Z_{24}, \right. \nonumber \\
& \left. Z_{3} Z_{5} Z_{6} Z_{8} Z_{13} Z_{15} Z_{17} Z_{23}, \right. \nonumber \\
& \left. Z_{4} Z_{5} Z_{8} Z_{11} Z_{14} Z_{16} Z_{18} Z_{20}, \right. \nonumber \\
& \left. Z_{1} Z_{5} Z_{10} Z_{12} Z_{14} Z_{17} Z_{19} Z_{22} \right. \nonumber \\
& \left. Z_{1} Z_{2} Z_{3} Z_{6} Z_{13} Z_{16} Z_{17} Z_{18} \right. \nonumber \\
& \left. Z_{6} Z_{7} Z_{10} Z_{11} Z_{15} Z_{18} Z_{19} Z_{24} \right \rangle \label{eq:24_10_4_code}
\end{flalign}

\begin{flalign}
S_{\nkd{24}{4}{5}} = & \left \langle X_{6} X_{9} X_{13} X_{18} X_{19} X_{20} X_{21} X_{22}, \right. \nonumber \\
& \left. X_{7} X_{11} X_{14} X_{17} X_{19} X_{21} X_{23} X_{24}, \right. \nonumber \\
& \left. X_{8} X_{10} X_{13} X_{15} X_{17} X_{20} X_{22} X_{23}, \right. \nonumber \\
& \left. X_{5} X_{12} X_{14} X_{16} X_{17} X_{18} X_{20} X_{24}, \right. \nonumber \\
& \left. X_{3} X_{12} X_{14} X_{15} X_{17} X_{19} X_{22} X_{24}, \right. \nonumber \\
& \left. X_{2} X_{7} X_{13} X_{14} X_{16} X_{17} X_{18} X_{19}, \right. \nonumber \\
& \left. X_{1} X_{11} X_{13} X_{15} X_{18} X_{19} X_{23} X_{24}, \right. \nonumber \\
& \left. X_{4} X_{10} X_{13} X_{16} X_{20} X_{21} X_{22} X_{24}, \right. \nonumber \\
& \left. X_{2} X_{6} X_{14} X_{16} X_{18} X_{21} X_{22} X_{23}, \right. \nonumber \\
& \left. X_{4} X_{5} X_{15} X_{16} X_{17} X_{18} X_{21} X_{22}, \right. \nonumber \\
& \left. Z_{1} Z_{3} Z_{6} Z_{7} Z_{8} Z_{11} Z_{19} Z_{23}, \right. \nonumber \\
& \left. Z_{2} Z_{4} Z_{5} Z_{6} Z_{9} Z_{11} Z_{18} Z_{21}, \right. \nonumber \\
& \left. Z_{3} Z_{5} Z_{7} Z_{10} Z_{11} Z_{12} Z_{17} Z_{24}, \right. \nonumber \\
& \left. Z_{4} Z_{6} Z_{8} Z_{9} Z_{10} Z_{12} Z_{20} Z_{22}, \right. \nonumber \\
& \left. Z_{2} Z_{3} Z_{4} Z_{5} Z_{6} Z_{10} Z_{16} Z_{22}, \right. \nonumber \\
& \left. Z_{1} Z_{4} Z_{6} Z_{7} Z_{9} Z_{10} Z_{13} Z_{21}, \right. \nonumber \\
& \left. Z_{1} Z_{2} Z_{5} Z_{6} Z_{7} Z_{12} Z_{14} Z_{18}, \right. \nonumber \\
& \left. Z_{1} Z_{3} Z_{4} Z_{8} Z_{11} Z_{12} Z_{15} Z_{24}, \right. \nonumber \\
& \left. Z_{1} Z_{2} Z_{8} Z_{9} Z_{10} Z_{11} Z_{13} Z_{23}, \right. \nonumber \\
& \left. Z_{1} Z_{3} Z_{5} Z_{8} Z_{9} Z_{10} Z_{15} Z_{20} \right \rangle
\end{flalign}

\begin{flalign}
S_{\nkd{42}{10}{6}} = & \left \langle X_{4} X_{12} X_{19} X_{26} X_{32} X_{33} X_{35} X_{40}, \right. \nonumber \\
& \left. X_{6} X_{16} X_{20} X_{27} X_{31} X_{40} X_{41} X_{42}, \right. \nonumber \\
& \left. X_{5} X_{15} X_{19} X_{26} X_{37} X_{39} X_{40} X_{41}, \right. \nonumber \\
& \left. X_{7} X_{10} X_{21} X_{28} X_{32} X_{34} X_{41} X_{42}, \right. \nonumber \\
& \left. X_{8} X_{11} X_{13} X_{22} X_{29} X_{33} X_{34} X_{42}, \right. \nonumber \\
& \left. X_{1} X_{3} X_{12} X_{22} X_{24} X_{34} X_{35} X_{38}, \right. \nonumber \\
& \left. X_{2} X_{14} X_{17} X_{22} X_{23} X_{30} X_{38} X_{39}, \right. \nonumber \\
& \left. X_{4} X_{9} X_{18} X_{25} X_{36} X_{38} X_{39} X_{40}, \right. \nonumber \\
& \left. X_{1} X_{6} X_{10} X_{26} X_{27} X_{28} X_{31} X_{38}, \right. \nonumber \\
& \left. X_{3} X_{14} X_{19} X_{23} X_{24} X_{29} X_{35} X_{41}, \right. \nonumber \\
& \left. X_{2} X_{9} X_{20} X_{23} X_{24} X_{25} X_{30} X_{42}, \right. \nonumber \\
& \left. X_{4} X_{15} X_{21} X_{23} X_{25} X_{26} X_{34} X_{36}, \right. \nonumber \\
& \left. X_{2} X_{11} X_{18} X_{25} X_{31} X_{32} X_{33} X_{39}, \right. \nonumber \\
& \left. X_{5} X_{13} X_{16} X_{22} X_{25} X_{26} X_{27} X_{37}, \right. \nonumber \\
& \left. X_{7} X_{11} X_{17} X_{27} X_{28} X_{29} X_{32} X_{39}, \right. \nonumber \\
& \left. X_{5} X_{14} X_{20} X_{27} X_{30} X_{33} X_{35} X_{41}, \right. \nonumber \\
& \left. Z_{5} Z_{6} Z_{7} Z_{14} Z_{20} Z_{27} Z_{30} Z_{41}, \right. \nonumber \\
& \left. Z_{7} Z_{10} Z_{11} Z_{12} Z_{21} Z_{28} Z_{32} Z_{34}, \right. \nonumber \\
& \left. Z_{6} Z_{10} Z_{11} Z_{16} Z_{20} Z_{27} Z_{31} Z_{42}, \right. \nonumber \\
& \left. Z_{8} Z_{11} Z_{12} Z_{13} Z_{14} Z_{22} Z_{29} Z_{33}, \right. \nonumber \\
& \left. Z_{1} Z_{3} Z_{9} Z_{12} Z_{14} Z_{24} Z_{35} Z_{38}, \right. \nonumber \\
& \left. Z_{2} Z_{9} Z_{14} Z_{15} Z_{17} Z_{23} Z_{30} Z_{39}, \right. \nonumber \\
& \left. Z_{4} Z_{9} Z_{15} Z_{16} Z_{18} Z_{25} Z_{36} Z_{40}, \right. \nonumber \\
& \left. Z_{5} Z_{10} Z_{15} Z_{16} Z_{19} Z_{26} Z_{37} Z_{41}, \right. \nonumber \\
& \left. Z_{7} Z_{11} Z_{17} Z_{18} Z_{19} Z_{29} Z_{32} Z_{39}, \right. \nonumber \\
& \left. Z_{2} Z_{9} Z_{13} Z_{20} Z_{21} Z_{25} Z_{30} Z_{42}, \right. \nonumber \\
& \left. Z_{1} Z_{4} Z_{13} Z_{15} Z_{21} Z_{26} Z_{34} Z_{36}, \right. \nonumber \\
& \left. Z_{1} Z_{5} Z_{13} Z_{16} Z_{17} Z_{22} Z_{27} Z_{37}, \right. \nonumber \\
& \left. Z_{4} Z_{5} Z_{6} Z_{12} Z_{19} Z_{26} Z_{35} Z_{40}, \right. \nonumber \\
& \left. Z_{1} Z_{6} Z_{10} Z_{17} Z_{18} Z_{28} Z_{31} Z_{38}, \right. \nonumber \\
& \left. Z_{3} Z_{14} Z_{19} Z_{20} Z_{21} Z_{23} Z_{35} Z_{41}, \right. \nonumber \\
& \left. Z_{6} Z_{7} Z_{8} Z_{9} Z_{21} Z_{28} Z_{36} Z_{42} \right \rangle
\end{flalign}

\onecolumngrid
\section{Circuit Diagrams}
\begin{figure}[!htbp]
\centering
\resizebox{\textwidth}{!}{\input{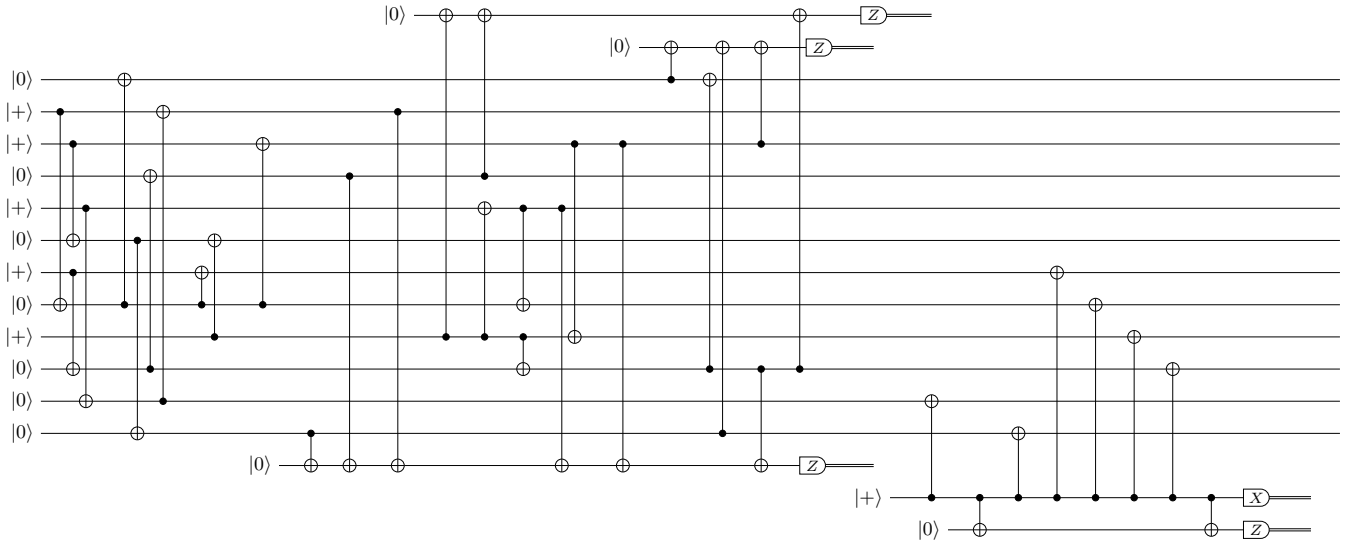}}
\caption{State preparation circuit for the $\nkd{12}{2}{4}$ Carbon code.
All single faults are detected, and all pairs of faults are either detected or result in errors at the output which cannot be mistaken for weight-1 errors, resulting in post-selection later in the QEC gadget.}
\label{fig:carbon_prep}
\end{figure}

\begin{figure}[!htbp]
\centering
\resizebox{\textwidth}{!}{\input{16_1_4_prep.tikz}}
\caption{State preparation circuit for the $\nkd{16}{1}{4}$ surface code.
All single faults are detected, and all pairs of faults are either detected or result in errors at the output which cannot be mistaken for weight-1 errors, resulting in post-selection later in the QEC gadget.}
\label{fig:16_1_4_prep}
\end{figure}

\begin{figure}[!htbp]
\centering
\resizebox{\textwidth}{!}{\input{16_2_4_prep.tikz}}
\caption{State preparation circuit for the $\nkd{16}{2}{4}$ colour code.
All single faults are detected, and all pairs of faults are either detected or result in errors at the output which cannot be mistaken for weight-1 errors, resulting in post-selection later in the QEC gadget.}
\label{fig:16_2_4_prep}
\end{figure}

\begin{figure}[!htbp]
\centering
\resizebox{\textwidth}{!}{\input{16_4_4_8_prep.tikz}}
\caption{State preparation circuit for the $\nkd{16}{4}{4}$ twisted colour code.
All single faults are detected, and all pairs of faults are either detected or result in errors at the output which cannot be mistaken for weight-1 errors, resulting in post-selection later in the QEC gadget.}
\label{fig:16_4_4_8_prep}
\end{figure}

\begin{figure}[!htbp]
\centering
\resizebox{\textwidth}{!}{\input{16_4_4_6_prep.tikz}}
\caption{State preparation circuit for the $\nkd{16}{4}{4}$ code with weight-6 stabilizers in \cref{eq:16_4_4_6_code}.
All single faults are detected, and all pairs of faults are either detected or result in errors at the output which cannot be mistaken for weight-1 errors, resulting in post-selection later in the QEC gadget.}
\label{fig:16_4_4_6_prep}
\end{figure}

\begin{figure}[!htbp]
\centering
\resizebox{\textwidth}{!}{\input{16_6_4_prep.tikz}}
\caption{State preparation circuit for the $\nkd{16}{6}{4}$ Reed-Muller code.
All single faults are detected, and all pairs of faults are either detected or result in errors at the output which cannot be mistaken for weight-1 errors, resulting in post-selection later in the QEC gadget.}
\label{fig:16_6_4_prep}
\end{figure}

\begin{figure}[!htbp]
\centering
\resizebox{\textwidth}{!}{\input{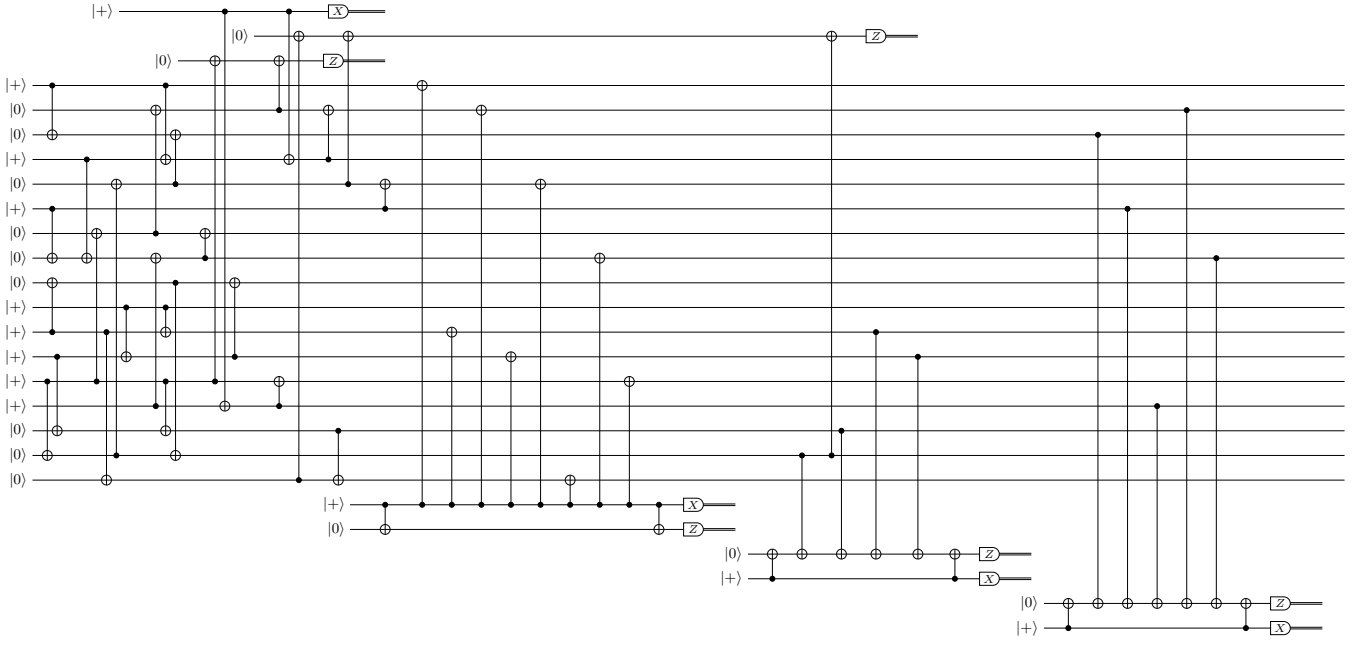}}
\caption{State preparation circuit for the $\nkd{17}{1}{5}$ colour code.
All faults or pairs of faults that propagate to errors at the output with weight greater than one or two, respectively, are detected.}
\label{fig:17_1_5_prep}
\end{figure}

\begin{turnpage}
\begin{figure}[!htbp]
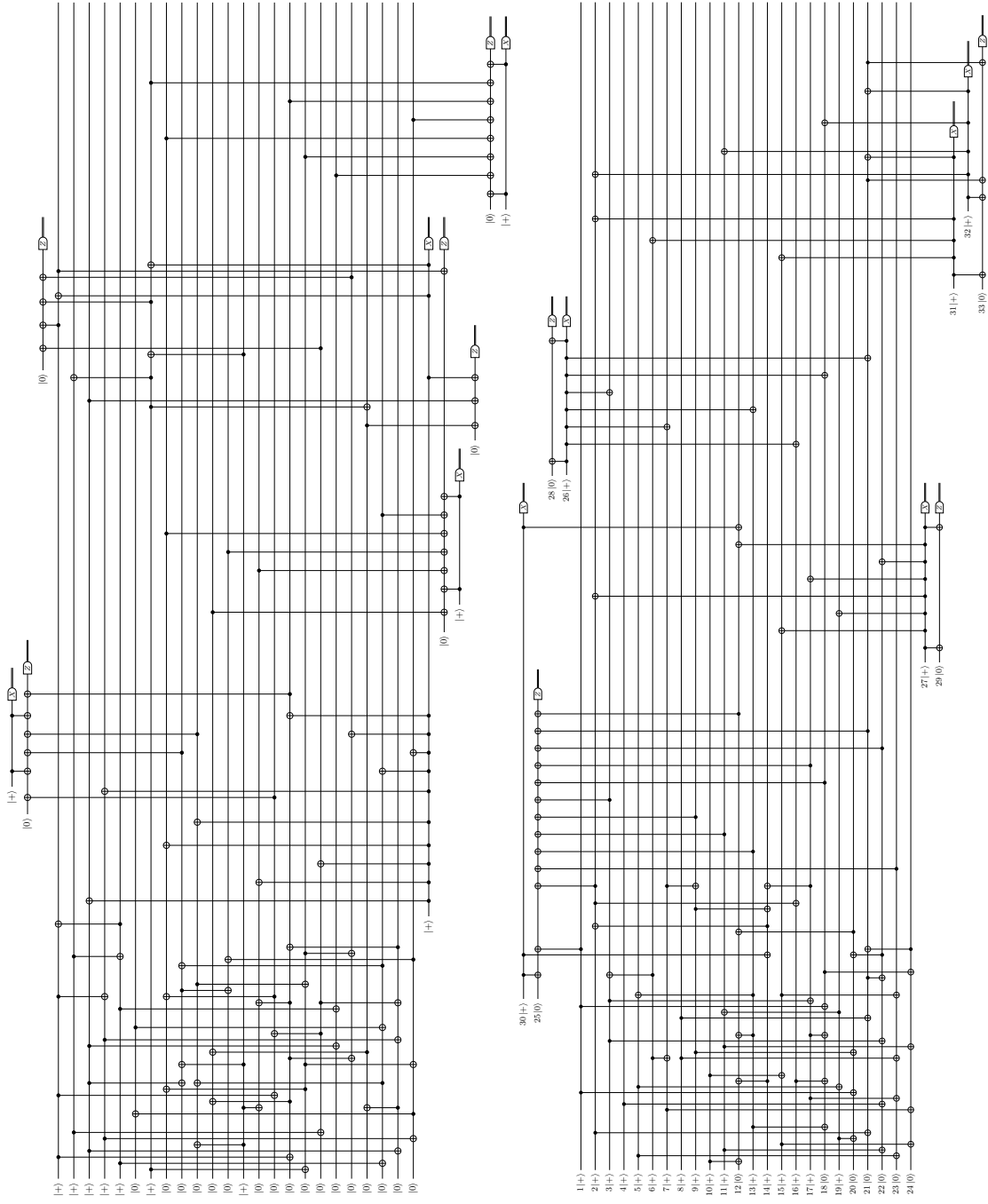

\centering
\resizebox{\textwidth}{!}{\input{24_10_4_prep.tikz}} \\ 
\resizebox{\textwidth}{!}{\input{24_10_4_plus_prep.tikz}}
\caption{State preparation circuits for the $\overline{\ket{0}^{10}}$ (top) and $\overline{\ket{+}^{10}}$ (bottom) states of the $\nkd{24}{10}{4}$ two-block group algebra code in \cref{eq:24_10_4_code}.
All single faults are detected, and all pairs of faults are either detected or result in errors at the output which cannot be mistaken for weight-1 errors, resulting in post-selection later in the QEC gadget.}
\label{fig:24_10_4_prep}
\end{figure}
\end{turnpage}

\begin{turnpage}
\begin{figure}[!htbp]
\centering
\resizebox{\textheight}{!}{\includegraphics[scale=1]{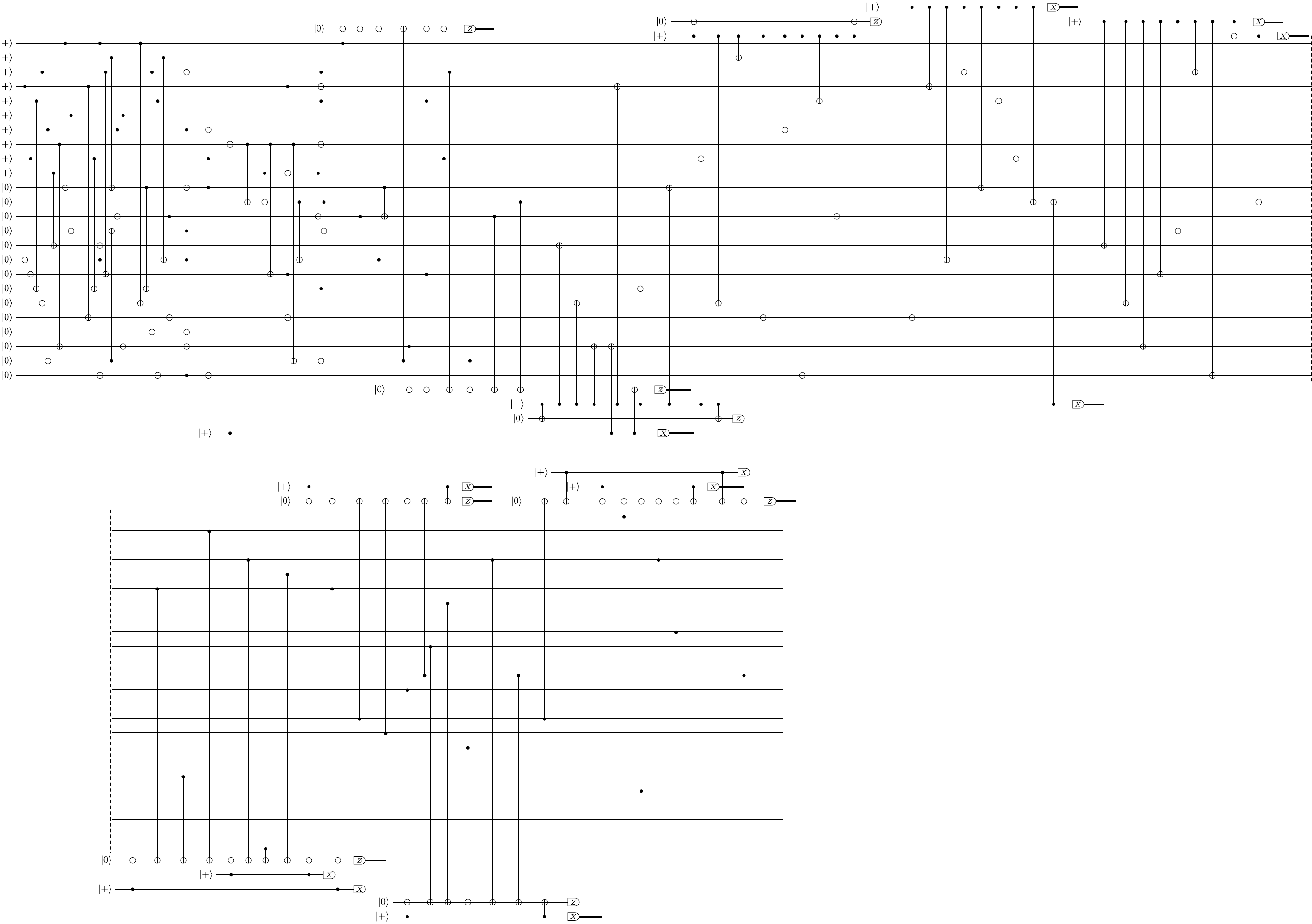}}
\caption{State preparation circuit for the $\nkd{24}{4}{5}$ two-block group algebra code.
All faults and pairs of faults that propagate to errors at the output with weight greater than one or two respectively are detected.
Vertical dotted line indicates a `cut' in the circuit, added to aid visibility.}
\label{fig:24_4_5_prep}
\end{figure}

\begin{figure}[!htbp]
\centering
\resizebox{\textheight}{!}{\input{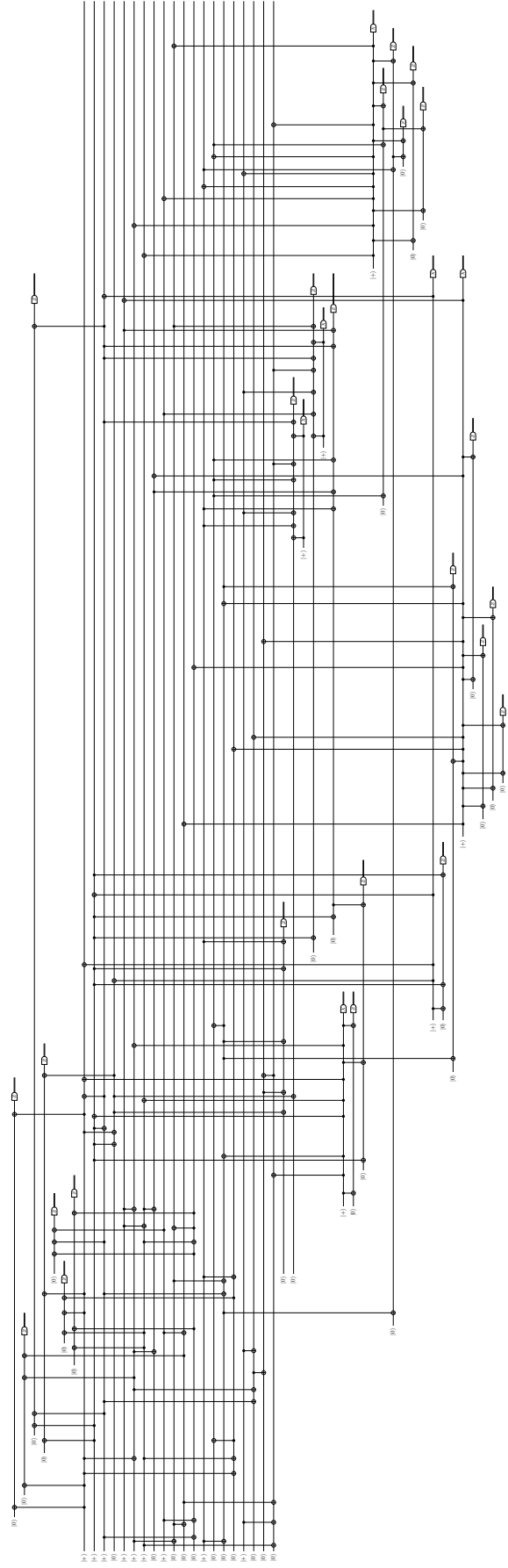}}
\caption{State preparation circuit for the $\nkd{20}{2}{6}$ Bravyi-Leemhuis-Terhal code.
All sets of faults up to size three that propagate to errors at the output with weight greater than their size are detected or result in post-selection later in the QEC gadget.}
\label{fig:20_2_6_prep}
\end{figure}

\begin{figure}[!htbp]
\centering
\resizebox{\textheight}{!}{\includegraphics[scale=1]{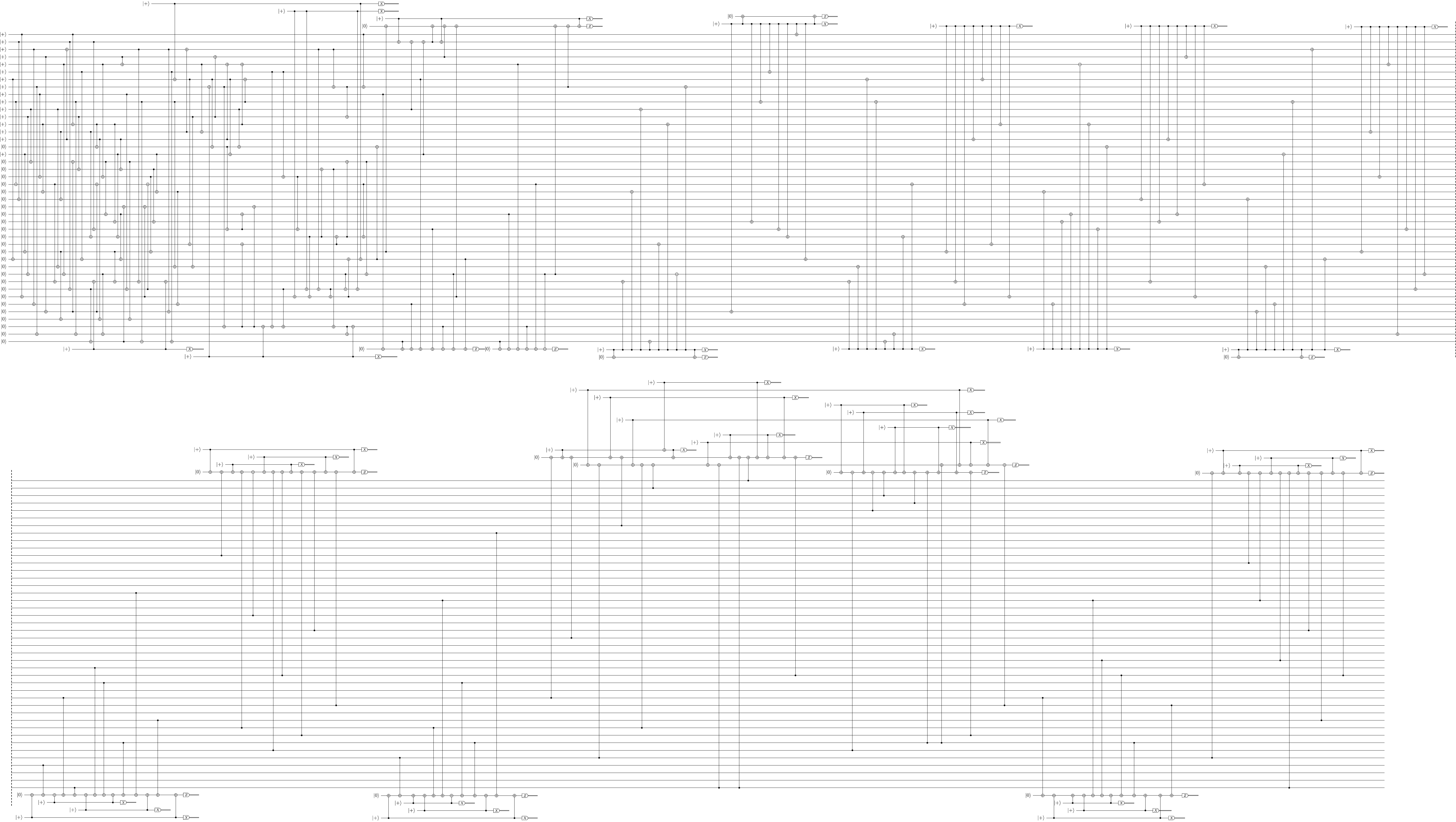}}
\caption{State preparation circuit for the $\nkd{42}{10}{6}$ two-block group algebra code.
All sets of faults up to size three that propagate to errors at the output with weight greater than their size are detected or result in post-selection later in the QEC gadget.
Vertical dotted line indicates a `cut' in the circuit, added to aid visibility.}
\label{fig:42_10_6_prep}
\end{figure}

\begin{figure}[!htbp]
\centering
\resizebox{\textheight}{!}{\includegraphics[scale=1]{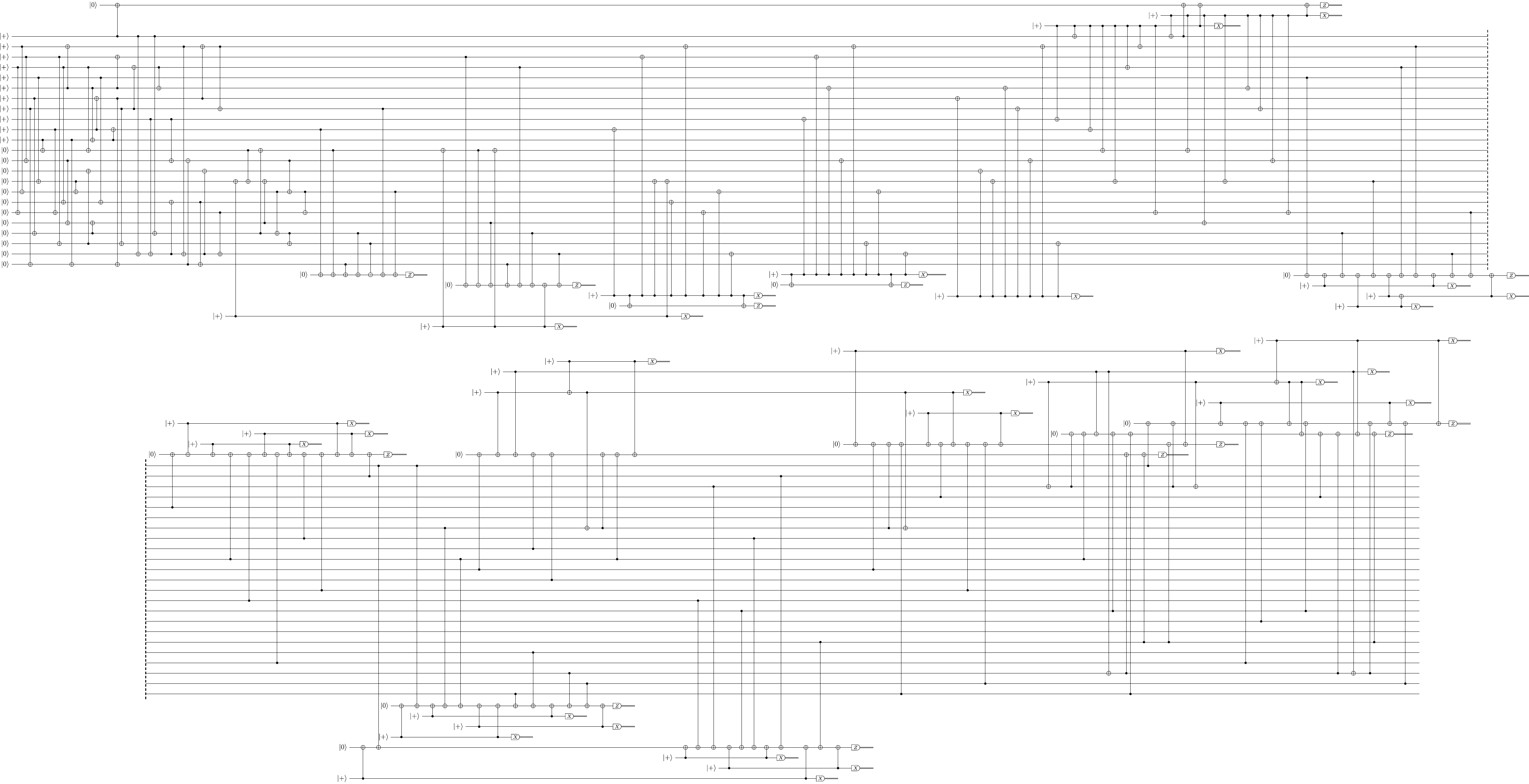}}
\caption{State preparation circuit for the $\nkd{23}{1}{7}$ Golay code.
All sets of faults up to size three that propagate to errors at the output with weight greater than their size are detected.}
\label{fig:23_1_7_prep}
\end{figure}
\end{turnpage}
\end{document}